\documentclass{article}

\usepackage[paper=letterpaper,hmargin=2.5cm,top=2.5cm,bottom=2.5cm,bindingoffset=0cm,twoside,hoffset=0pt,marginparsep=0pt,marginparwidth=0pt]{geometry}

\usepackage{hhline}
\usepackage{amssymb}
\usepackage{amsmath}
\usepackage{amsfonts}
\usepackage{array}
\usepackage{graphicx}
\usepackage[dvipsnames]{xcolor}
\usepackage{xspace}
\usepackage{ulem}
\usepackage{lscape}
\usepackage{wasysym}
\usepackage{xcolor}
\usepackage{authblk}

\usepackage{multirow,rotating}
\usepackage{dsfont}

\usepackage{tikz}
\usetikzlibrary{matrix}

\def\gsim{\raise0.3ex\hbox{$\;>$\kern-0.75em\raise-1.1ex\hbox{$\sim\;$}}}
\def\lsim{\raise0.3ex\hbox{$\;<$\kern-0.75em\raise-1.1ex\hbox{$\sim\;$}}}

\newcommand{\qedqcd}{\mbox{QED}\otimes\mbox{QCD}}

\newcommand{\ba}[1]{\begin{eqnarray} \label{(#1)}}
\newcommand{\ea}{\end{eqnarray}}

\newcommand{\seq}{\begin{subequations}}
\newcommand{\sen}{\end{subequations}}
\newcommand{\eq}{\begin{eqnarray}}
\newcommand{\en}{\end{eqnarray}}

\newcommand{\AddrSAPHIR}{
Millenium Institute for Subatomic Physics at High-Energy Frontier (SAPHIR)\\ 
Fern\'andez Concha 700, Santiago,
Chile
}

\newcommand{\AddrUNAB}{
Departamento de Ciencias F\'isicas
Universidad Andres Bello, 
Sazi\'e 2212, Santiago, Chile
}

\newcommand{\AddrUTFSM}{
Universidad T\'ecnica Federico Santa Mar\'\i a,
Departamento de F\'isica\\
\mbox{Casilla 110-V}, Valpara\'\i so,  Chile}

\newcommand{\AddrCCTVal}{
Centro Científico-Tecnológico de
Valpara\'iso, Casilla
110-V, Valpara\'iso,
Chile
}

\newcommand{\AddrIAI}{
Instituto de Alta Investigaci\'{o}n\\
Universidad de Tarapac\'{a}, Casilla 7D, Arica, Chile
}

\def\gsim{\raise0.3ex\hbox{$\;>$\kern-0.75em\raise-1.1ex\hbox{$\sim\;$}}}
\def\lsim{\raise0.3ex\hbox{$\;<$\kern-0.75em\raise-1.1ex\hbox{$\sim\;$}}}

\title{RGE effects on the LFV scale from meson decays}

\author[1,2]{Marcela Gonz\'alez}
\author[2,3,4]{Sergey Kovalenko}
\author[5]{Nicol\'as A. Neill}
\author[1,2]{Jonatan Vignatti}

\affil[1]{{\textit{\small \AddrUTFSM}}}
\affil[2]{{\textit{\small \AddrCCTVal}}}
\affil[3]{{\textit{\small \AddrUNAB}}}
\affil[4]{{\textit{\small \AddrSAPHIR}}}
\affil[5]{{\textit{\small \AddrIAI}}}

\begin{document} 

\maketitle
\begin{abstract}
We consider the lepton-flavor violating (LFV) lepton-quark dimension-6 operators and analyze their contributions to the LFV leptonic decays of vector, pseudoscalar, and scalar neutral mesons 
$M\to \ell_1 \ell_2$ as well as to $\mu(\tau) \rightarrow \ell ee, \ell \gamma\gamma$ decays. These operators contribute to the purely leptonic processes via quark loop. On the basis of quark-hadron duality, we relate these loops to the appropriate meson-exchange contributions.
In this way, we extract lower bounds on the individual scales of  the studied LFV operators from the experimental and phenomenological limits on the leptonic decays of mesons and leptons. As a byproduct, we shall obtain new limits on the LFV leptonic decays of flavored mesons from the experimental bounds on the three-body lepton decays. 
We study the effects of QED and QCD radiative corrections to the LFV lepton-quark operators in question. We derive for them the one-loop matrix of the RGE evolution and examine its effect on the previously derived tree-level limits on these operators.
We show that the QED corrections are particularly 
relevant due to operator mixing. Specifically, for some of them the limits on their individual LFV scales improve by up to 3 orders of magnitude.
\end{abstract}

\section{Introduction}

In the Standard Model (SM), without the inclusion of neutrino masses, lepton flavor is an exactly conserved quantity.
On the contrary, neutrino oscillation experiments demonstrate  that neutrinos have small masses
and mix with each other, breaking lepton flavor conservation.
At the same time, lepton-flavor violating (LFV) processes in the sector of the charged leptons have not yet been observed experimentally. In principle, LFV must anyway leak from the neutrino sector to the charged lepton one via radiative corrections. However, this effect is highly suppressed by the small neutrino mass squared differences $\Delta m^{2}_{\nu}$
at a level far beyond the realistic experimental sensitivities. 
On the other hand, 
the charged lepton sector can acquire other LFV contributions from the physics beyond SM residing at a certain  high-energy scale  $\Lambda$.
Consequently, searches for 
charged LFV are an important probe of physics beyond the SM.

Many experiments have searched and put bounds on charged LFV interactions, e.g., in the processes like $\ell \to e \gamma$ \cite{TheMEG:2016wtm,Aubert:2009ag}, $\ell^- \to e^- e^+ e^-$ \cite{Hayasaka:2010np,Bellgardt:1987du} (with $\ell=\mu,\tau$), $\mu^--e^-$ conversion in nuclei \cite{Dohmen:1993mp}, and rare meson decays.
There are also collider searches looking for LFV in the scalar sector (e.g., $H\to \tau^\pm \mu^\mp$) \cite{Aad:2019ugc,Sirunyan:2019shc}, in $Z$-boson decays \cite{CMS:2015hga,Aad:2020gkd}, or in decays of heavy resonances \cite{Sirunyan:2018zhy,Aaboud:2018jff}. 
Additionally, future experiments, such as Mu3e \cite{Arndt:2020obb}, COMET \cite{Angelique:2018svf,Adamov:2018vin}, Mu2e \cite{Miscetti:2020gkk}, and MEG-II \cite{Baldini:2018nnn} are expected to tighten current limits from 1 and up to 4 orders of magnitude, depending on the process.

In this work, we study LFV  
in a model-independent way in the context of an effective  LFV quark-lepton Lagrangian with the operators up to dimension-6, which can contribute to various LFV processes such as nuclear $\mu-e$-conversion \cite{Faessler:2004jt,Faessler:2004ea,Faessler:2005hx,Dib:2018rpy}, deep inelastic 
$\ell_{1}-\ell_{2}$-conversion \cite{Gninenko:2018num}, 
decays of mesons 
$M\to \ell_1 \ell_2$  and leptons $\mu(\tau) \rightarrow \ell ee, \ell \gamma\gamma$ \cite{Dib:2018rpy}. Previously, for the quark flavor-diagonal case, the Wilson Coefficients (WC) and the individual LFV scales for these operators have been constrained in the literature from the existing experimental limits on these processes (for a summary, see for example Ref.~\cite{Dib:2018rpy}).
We include in the analysis quark flavor non-diagonal operators, for which we derive new tree-level limits from current limits on the LFV rates of flavored mesons.
Here we examine the effect of QED and QCD corrections on the LFV dimension-6 quark-lepton operators.  
We find that QED corrections are particularly relevant due to operator mixing. This effect in some cases significantly improves the above-mentioned tree-level constraints on the WCs. 

The paper is organized as follows. In section \ref{sec:Limits-on-Mesons}, using unitary-based arguments, we derive indirect bounds on the LFV meson decays $M\to \ell_1 \ell_2$ from the experimental limits on $\mu(\tau) \rightarrow \ell ee, \ell \gamma\gamma$ decays. 
In section \ref{sec:effectiveL}, we describe the effective quark-lepton and meson-lepton LFV Lagrangians. Matching these Lagrangians on mass-shell, in section \ref{sec:limitsnocorr}, we relate the meson couplings with the quark-lepton Wilson Coefficients and extract tree-level limits for them from the leptonic meson decays. 
In section \ref{sec:limitswithcorr}, we derive the 1-loop QED and QCD corrections to the LFV operators and calculate the corresponding renormalization group evolution (RGE) matrix. Using this matrix, we find in section \ref{section:RGE_induced_limits} 
new QCD and QED improved lower limits 
on the individual LFV scales of these operators. 
Finally, in section \ref{sec:conclusions}, we outline our summary and conclusions.

\section{Limits on leptonic decays of mesons}
\label{sec:Limits-on-Mesons}
We start with the derivation of  indirect limits on the leptonic LFV decay rates of mesons 
$M\to \ell_1 \ell_2$ from the experimental limits on three-body LFV leptonic decays of $\mu$ and $\tau$,
\begin{align}
\rm{Br}(\mu^- \to e^- e^+ e^-) < &\, 1.0\times 10^{-12},\label{eq:l1tol2eee1}\\
\rm{Br}(\mu^- \to e^- \gamma\gamma) < &\, 7.2\times 10^{-11},\\
\rm{Br}(\tau^- \to e^- e^+ e^-) < &\, 2.7\times 10^{-8},\\
\rm{Br}(\tau^- \to \mu^- e^+ e^-) < &\, 1.8\times 10^{-8}.\label{eq:l1tol2eee4}
\end{align}
This can be done on the basis of the unitarity arguments as proposed in  Ref.~\cite{Nussinov:2000nm}. 
Previously this approach has been applied to the unflavored neutral mesons \cite{Nussinov:2000nm,Dib:2018rpy}. In this work, we extend this analysis and include the flavored pseudoscalars mesons $K^0,D^0,B^0,B_s^0$, and vector mesons $K^*,D^*,B^*,B_s^*$.
This will allows us to put bounds not only on the lepton branching ratios of these mesons but also on the quark-lepton dimension-6 flavor non-diagonal operators specified in the next section.

In order to constrain the partial widths $V\to \ell_1 \ell_2$ ($M=V,P,S$) of the flavored vector ($V$), pseudoscalar ($P$) and scalar ($S$) mesons from the three-body LFV leptonic decays in Eqs.~(\ref{eq:l1tol2eee1})-(\ref{eq:l1tol2eee4}), we use the  ``unitarity inspired'' arguments of Ref.~\cite{Nussinov:2000nm} leading to the following relations \cite{Nussinov:2000nm,Dib:2018rpy} between these quantities
\begin{eqnarray}\label{eq:brmuto3e}   
{\rm Br}(\mu \to 3 e) &=& 
\frac{\Gamma(V \to \mu  e) \, 
\Gamma(V \to e^+e^-)}{\Gamma(W \to e \bar\nu_e)^2} \, 
\biggl(\frac{M_W}{M_V}\biggr)^6 \,,
\end{eqnarray}

\begin{eqnarray}\label{eq:TauLEE-1}     
{\rm Br}(\tau^{-}\rightarrow \ell^{-} e^{+}e^{-}) &=&  
 \frac{\Gamma(V \to \tau \ell) \, 
\Gamma(V \to e^+e^-)}{\Gamma(W \to e \bar\nu_e)^2} \,                 
 \frac{\Gamma(\mu^{-}\to e^{-} \bar\nu_{e} \nu_{\mu})}{\Gamma(\tau \to All)}
  \left(\frac{M_W}{M_V}\right)^6   
\left(\frac{M_{\tau}}{M_{\mu}}\right)^5\,, 
\end{eqnarray}
where $M_V>m_\tau$ in the latter equation.
For pseudoscalar (scalar) mesons we have
\begin{eqnarray}\label{eq:MU-EGG-1} 
{\rm Br}(\mu^{-}\rightarrow e^{-} \gamma\gamma) &\approx&
\frac{\Gamma(P(S) \to \mu  e) \,
\Gamma(P (S)\to \gamma\gamma)}{\Gamma(W \to e \bar\nu_e)^2} \,
\left(\frac{M_W}{M_{P(S)}}\right)^6 \left(\frac{M_{\mu}}{2 M_{P(S)}}\right)^{4}.
\end{eqnarray}
For the vector mesons where the branching fraction $\mbox{Br}(V\to ee)$ has not been measured, we use the theoretical prediction 
\begin{equation}\label{eq:Vtoee}
\Gamma(V\to e^+ e^-) = \frac{4\pi \alpha_{em}^2 Q_q^2}{3} \frac{F_V^2}{m_V},
\end{equation}
with the meson decay constants $F_{V}$ quoted in Table \ref{tab:decayconstants}.
Our results for the flavored mesons, when they provide more stringent limits in comparison with the existing ones, are shown in Table \ref{tab:existingbounds}, which shows  for each meson the most stringent limit to our knowledge.
These values will be used in the next section to constrain the dimension-6 LFV effective operators.

\begin{figure}[htb]
\centering
\includegraphics[scale=0.9]{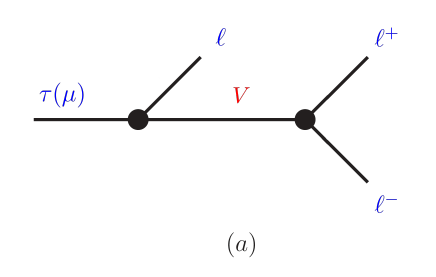}
\includegraphics[scale=0.9]{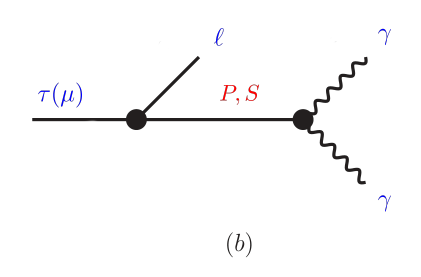}
\noindent
\caption{Vector ($V$), scalar ($S$) and pseudoscalar ($P$) meson contributions to the tree-body LFV decays:  
(a) $\tau(\mu) \to 3 \ell$ and 
(b) $\tau(\mu) \to \ell \, \gamma\gamma$.
\label{fig:tau3ell}}
\end{figure}

Let us point out once again that the relations (\ref{eq:brmuto3e})-(\ref{eq:MU-EGG-1})  are derived under the assumption of the dominance of a particular meson in the intermediate state of   leptonic processes
$\ell_{1}^- \to \ell_{2}^- e^+ e^-$ and $\mu^- \to e^- \gamma\gamma$, shown in the diagrams in 
Fig.~\ref{fig:tau3ell}. 
This assumption is supported by the analysis of  Ref.~\cite{Nussinov:2000nm} 
demonstrating that a strong cancellation of the contributions in Fig.~\ref{fig:tau3ell} with the contributions of other possible intermediate states is unlikely.

\begin{table}
\centering
\begin{tabular}{c|c||c|c}
Mode & Existing limit & Mode & Existing limit\\
\hline
\hline
\multicolumn{2}{c||}{from $\mu^- \to e^-\gamma\gamma$} & \multicolumn{2}{c}{from $\mu^- \to e^-e^+e^-$}\\
\hline
\hline
$\pi^0 \to \mu^\pm e^\mp$ & $5.8\times 10^{-11}$ \cite{Dib:2018rpy} & $\Upsilon \to \mu^\pm e^\mp$ & $2\times 10^{-9}$ \cite{Nussinov:2000nm}\\
\hline
$\eta \to \mu^\pm e^\mp$ & $6.2\times 10^{-9}$ \cite{Dib:2018rpy} & $K^* \to \mu^\pm e^\mp$ & $1.6\times 10^{-18}$ \\
\hline
$\eta' \to \mu^\pm e^\mp$ & $1.3\times 10^{-9}$ \cite{Dib:2018rpy} & $D^* \to \mu^\pm e^\mp$ & $4.9\times 10^{-18}\mbox{ GeV}/\Gamma_{D^*}$\\
\hline
$\eta_c \to \mu^\pm e^\mp$ & $5.9\times 10^{-7}$ \cite{Dib:2018rpy} & $B^* \to \mu^\pm e^\mp$ & $2.7\times 10^{-14}\mbox{ GeV}/\Gamma_{B^*}$\\
\hline
$f_0(500) \to \mu^\pm e^\mp$ & $1.6\times 10^{-15}$ \cite{Dib:2018rpy} & $B^*_s \to \mu^\pm e^\mp$ & $2.0\times 10^{-14}\mbox{ GeV}/\Gamma_{B^*_s}$\\
\hhline{--||==}
$f_0(980) \to \mu^\pm e^\mp$ & $1.0\times 10^{-10}$ \cite{Dib:2018rpy} & \multicolumn{2}{c}{from $\tau^- \to \mu^-e^+e^-$}\\
\hhline{--||==}
$a_0(980) \to \mu^\pm e^\mp$ & $6.2\times 10^{-11}$ \cite{Dib:2018rpy} & $J/\psi \to \tau^\pm \mu^\mp$ & $3.3\times 10^{-8}$ \cite{Dib:2018rpy} \\
\hline
$\chi_{c0}(1P) \to \mu^\pm e^\mp$ & $1.5\times 10^{-5}$ \cite{Dib:2018rpy} & $\Upsilon \to \tau^\pm \mu^\mp$ & $2.0\times 10^{-4}$ \cite{Dib:2018rpy}\\
\hline
$K^0_L \to \mu^\pm e^\mp$ & $4.7\times 10^{-12}$ & $K^* \to \tau^\pm \mu^\mp$ & $1.7\times 10^{-13}$\\
\hhline{==||--}
\multicolumn{2}{c||}{from $\mu^- - e^-$ conversion} & $D^* \to \tau^\pm \mu^\mp$ & $5.0\times 10^{-13}/\Gamma_{D^*}$\\
\hhline{==||--}
$\rho^0 \to \mu^\pm e^\mp$ & $3.5\times 10^{-24}$ \cite{Gutsche:2011bi} & $B^* \to \tau^\pm \mu^\mp$ & $2.7\times 10^{-9}/\Gamma_{B^*}$\\
\hline
$\omega \to \mu^\pm e^\mp$ & $6.2\times 10^{-27}$ \cite{Gutsche:2011bi} & $B_s^* \to \tau^\pm \mu^\mp$ & $2.0\times 10^{-9}/\Gamma_{B_s^*}$\\
\hline
$\phi \to \mu^\pm e^\mp$ & $1.3\times 10^{-21}$ \cite{Gutsche:2011bi} & \multicolumn{2}{c}{ } \\
\hhline{--||==}
$J/\psi \to \mu^\pm e^\mp$ & $3.5\times 10^{-13}$ \cite{Gutsche:2011bi} & \multicolumn{2}{c}{Direct limits}\\
\hhline{==||==}
\multicolumn{2}{c||}{from $\tau^- \to e^-e^+e^-$} & $K_L^0 \to \mu^\pm e^\mp$ & $4.7\times 10^{-12}$ \cite{Ambrose:1998us}\\
\hhline{==||--}
$J/\psi \to \tau^\pm e^\mp$ & $4.9\times 10^{-8}$ \cite{Dib:2018rpy} & $D^0 \to \mu^\pm e^\mp$ & $1.3\times 10^{-8}$ \cite{Aaij:2015qmj}\\
\hline
$\Upsilon \to \tau^\pm e^\mp$ & $2.9\times 10^{-4}$ \cite{Dib:2018rpy} & $B^0 \to \mu^\pm e^\mp$ & $1.0\times 10^{-9}$ \cite{Aaij:2017cza}\\
\hline
$K^* \to \tau^\pm e^\mp$ & $2.5\times 10^{-13}$ & $B^0_s \to \mu^\pm e^\mp$ & $5.4\times 10^{-9}$ \cite{Aaij:2017cza}\\
\hline
$D^* \to \tau^\pm e^\mp$ & $7.5\times 10^{-13}/\Gamma_{D^*}$ & $B^0 \to \tau^\pm e^\mp$ & $2.8\times 10^{-5}$ \cite{Aubert:2008cu}\\
\hline
$B^* \to \tau^\pm e^\mp$ & $4.0\times 10^{-9}/\Gamma_{B^*}$ & $B^0_s \to \tau^\pm e^\mp$ & $4.2\times 10^{-5}$ \cite{Aaij:2019okb}\\
\hline
$B_s^* \to \tau^\pm e^\mp$ & $3.0\times 10^{-9}/\Gamma_{B_s^*}$ & $B^0 \to \tau^\pm \mu^\mp$ & $1.4\times 10^{-5}$ \cite{Aaij:2019okb}\\
\hline
\hline
\end{tabular}\caption{Upper bounds for the branching ratios of two-body LFV decays of mesons. The existing limits are supplied with the corresponding reference, while new limits derived in section \ref{sec:Limits-on-Mesons} are not. 
When there is more than one bound in the literature, we quote the most stringent limit to our knowledge.
The limits for the vector mesons $K^*$, $D^*$, $B^*$, and $B^*_s$ were  derived from Eqs.~(\ref{eq:brmuto3e}) and (\ref{eq:TauLEE-1}). See a comment about their viability at the end of 
Sec.~\ref{sec:Limits-on-Mesons} 
}\label{tab:existingbounds}
\end{table}

\section{LFV effective Lagrangians}\label{sec:effectiveL}

Let us consider the effective quark-lepton LFV Lagrangian with the operators up to dimension 6. It reads
\cite{Dib:2018rpy}
\begin{eqnarray}
\label{eq:l-q-operators-2}
\mathcal{L}_{\ell q} &=& \frac{1}{\Lambda^2} \, \biggl[
\ \bar{\ell}_{1}\, 
\Big(C^{(q_1 q_2)SS}_{\ell_{1}\ell_{2}} \,+\, 
 C^{(q_1 q_2)PS}_{\ell_{1}\ell_{2}} \gamma^5\Big) \, \ell_{2}  \cdot  
\bar{q_1}    q_2
\,+\, 
\bar{\ell}_{1}\, 
\Big(C^{(q_1 q_2)SP}_{\ell_{1}\ell_{2}} \,+\, 
 C^{(q_1 q_2)PP}_{\ell_{1}\ell_{2}} \gamma^5\Big)\, \ell_{2}  \cdot  
\bar{q_1} \gamma_5  q_2
\nonumber\\
&+&
\bar{\ell}_{1}\,  
\Big(C^{(q_1 q_2)VV}_{\ell_{1}\ell_{2}} \, \gamma^\mu \,+\, 
 C^{(q_1 q_2)AV}_{\ell_{1}\ell_{2}} \, \gamma^\mu\gamma^5\Big)\, \ell_{2} \cdot  
\bar{q_1} \gamma_\mu q_2 \,+\, 
\bar{\ell}_{1}\,  
\Big(C^{(q_1 q_2)VA}_{\ell_{1}\ell_{2}} \, \gamma^\mu \,+\, 
 C^{(q_1 q_2)AA}_{\ell_{1}\ell_{2}} \, \gamma^\mu\gamma^5\Big)\, \ell_{2} \cdot  
\bar{q_1} \gamma_\mu\gamma_5 q_2 
\nonumber\\ 
\label{eq:T-term-1}
&+&
C^{(q_1 q_2)TT}_{\ell_{1}\ell_{2}} \, 
\bar{\ell}_{1} \sigma^{\mu\nu} \ell_{2} \cdot 
\bar{q_1} \sigma_{\mu\nu} q_2 \ \biggr]
 + {\rm H.c.},
\end{eqnarray}
where $\ell = e,\mu,\tau$ and $\Lambda$ is a scale of an underlying LFV physics. The dimensionless parameters  
${C}^{(q)AB}_{\ell_{1}\ell_{2}}$ are Wilson Coefficients (WC) in the operator basis with definite P-parity.
The WCs are free parameters to be determined from the experiment. They can also be related to more fundamental parameters of the underlying LFV ultraviolet (UV) completions  at the cutoff scale $\Lambda$.  So that, having limits on the WCs, one can constrain UV models reducible at low energies to the effective operators in Eq.~(\ref{eq:l-q-operators-2}).

These effective operators contribute to the LFV leptonic decays of neutral mesons, $M\to \ell_1 \ell_2$, 
specifically the vector mesons $\rho,\omega,\phi,J/\psi,\Upsilon,K^{*},D^{*},B^{*},B_s^{*}$,
the pseudoscalar and scalar mesons $\pi^0$, $\eta$, $\eta'$, $\eta_c$, $f_0(500)$, $f_0(980)$, $a_0(980)$, $\chi_{c0}(1P)$, $K^0$, $D^0$, $B^0$, $B^0_s$,
as well as to the nuclear $\mu-e$-conversion \cite{Faessler:2004jt,Faessler:2004ea,Faessler:2005hx,Gutsche:2011bi} and deep inelastic $\ell_{1}-\ell_{2}$-conversion \cite{Gninenko:2018num}.
At the quark loop level, they also contribute to purely leptonic processes such as $\mu(\tau) \rightarrow \ell ee, \ell \gamma\gamma$ \cite{Dib:2018rpy,Nussinov:2000nm}.
These loop contributions are related through the quark-hadron duality to the sum over all tree-level contributions mediated by mesons with the allowed quantum numbers \cite{Nussinov:2000nm}.
The relevant meson-lepton operators that contribute to $\ell \to \ell' e^+ e^-$ are
\begin{eqnarray}
\label{eq:L-meson-lepton}
\mathcal{L}_{\ell M} &=& 
V_{\mu} \left(g^{(V)}_{V\ell_{1}\ell_{2}} \, 
[\bar{\ell}_{1} \gamma^{\mu} \ell_{2}]+
g_{V\ell_{1}\ell_{2}}^{(A)} \, 
[\bar{\ell}_{1} 
\gamma^{\mu}\gamma_{5} \ell_{2} ] \right) +
A_{\mu} \left(g^{(V)}_{A\ell_{1}\ell_{2}} \, 
[ \bar{\ell}_{1} \gamma^{\mu} \ell_{2} ]+
g_{A\ell_{1}\ell_{2}}^{(A)} \,
[ \bar{\ell}_{1} 
\gamma^{\mu}\gamma_{5} \ell_{2} ] \right) 
\nonumber\\
&+&
\frac{g^{(T)}_{V\ell_1\ell_2}}{M_V} \, 
F^{V}_{\mu\nu}\, \left[\bar{\ell}_{1} \sigma^{\mu\nu} \ell_{2}\right]  
+\frac{g^{(T)}_{A\ell_1\ell_2}}{M_A} \, 
F^{A}_{\mu\nu}\, \left[\bar{\ell}_{1} \sigma^{\mu\nu} \gamma^5 \ell_{2}\right]
\nonumber\\
&+&S\, \left(g^{(S)}_{S\ell_{1}\ell_{2}} \,
[ \bar{\ell}_{1}  \ell_{2} ]
+ 
g_{S\ell_{1}\ell_{2}}^{(P)} \, 
[ \bar{\ell}_{1} \gamma_{5} \ell_{2}] \right) + 
P\, \left( i g^{(S)}_{P\ell_{1}\ell_{2}} \, [ \bar{\ell}_{1}   \ell_{2} ] + 
i g_{P\ell_{1}\ell_{2}}^{(P)} \, [\bar{\ell}_{1} \gamma_{5} \ell_{2} ]
\right) 
\nonumber\\
&+&
\frac{\partial_\mu P}{M_P} \, 
\left(g^{(V)}_{P\ell_1\ell_2} [ \bar{\ell}_{1} \gamma^\mu \ell_{2} ] +
g^{(A)}_{P\ell_1\ell_2}  [ \bar{\ell}_{1} \gamma^\mu\gamma^5 \ell_{2} ] \right) 
\,+\, {\rm H.c.},
\end{eqnarray}
where $V$, $A$, $P$, and $S$ represent vector, axial, pseudoscalar, and scalar meson fields, respectively, and \mbox{$F_{\mu\nu}^{V(A)}=\partial_\mu V(A)_\nu - \partial_\nu V(A)_\mu$}.
The leptonic decay rates of all these mesons $M\to \ell_1 \ell_2$ can be calculated from the \mbox{Lagrangian} (\ref{eq:L-meson-lepton}) and expressed in terms of the effective coupling constants $g^{(M)}_{AB}$. The corresponding expressions are given in Appendix \ref{App:M-rates}.
In the next section, we relate these coupling constants with the appropriate WCs of the Lagrangian (\ref{eq:l-q-operators-2}).

\section{Tree-level limits on the Wilson Coefficients}\label{sec:limitsnocorr}

The relation between quark-lepton, $C^{(q_1 q_2)AB}_{\ell_1 \ell_2}$, and meson-lepton, $g_{M}$, couplings from Eqs.~(\ref{eq:l-q-operators-2}) and (\ref{eq:L-meson-lepton}), respectively, can be derived from the on-shell matching condition 
\cite{Faessler:2004jt,Faessler:2004ea,Faessler:2005hx}
\begin{equation}\label{matching} 
\langle \ell_{1}^+ \, \ell_{2}^-|{\cal L}_{eff}^{lq}|M\rangle \approx 
\langle \ell_{1}^+ \, \ell_{2}^-|{\cal L}_{eff}^{lM}|M \rangle.  
\end{equation}
These relations for unflavored mesons were derived in Ref.~\cite{Dib:2018rpy}. We extended this analysis to include flavored mesons. The resulting complete list of the meson-lepton couplings in terms of the WCs reads

\vspace{2mm}
\noindent
\underline{Vector mesons:}
\begin{eqnarray}
g^{(V/A)}_{\rho^0 \ell_1\ell_2} &=& \frac{M_\rho^2}{\Lambda^2} 
\, f_{\rho}  \, C^{(3)VV/AV}_{\ell_1\ell_2}
\,,\quad 
g^{(V/A)}_{\omega \ell_1\ell_2} \, = \,  \frac{3 M_\omega^2}{\Lambda^2} 
\, f_{\omega}  \, C^{(0)VV/AV}_{\ell_1\ell_2}
\,,\quad 
g^{(V/A)}_{\phi \ell_1\ell_2} \, = \,  - \frac{3 M_\phi^2}{\Lambda^2} 
\, f_{\phi}  \, C^{(s) VV/AV}_{\ell_1\ell_2}
\,,\nonumber\\ 
g^{(V/A)}_{J/\psi \ell_1\ell_2} &=& \frac{M_{J/\psi}^2}{\Lambda^2} 
\, f_{J/\psi}  \, C^{(c) VV/AV}_{\ell_1\ell_2}
\,,\quad 
g^{(V/A)}_{\Upsilon \ell_1\ell_2} \, = \,  \frac{M_{\Upsilon}^2}{\Lambda^2} 
\, f_{\Upsilon}  \, C^{(b) VV/AV}_{\ell_1\ell_2}
\,,\nonumber
g^{(V/A)}_{K^* \ell_1\ell_2} \, = \,  \frac{M_{K^*}^2}{\Lambda^2} 
\, f_{K^*}  \, C^{(ds) VV/AV}_{\ell_1\ell_2}
\,,\nonumber\\ 
g^{(V/A)}_{D^* \ell_1\ell_2} &=& \frac{M_{D^*}^2}{\Lambda^2} 
\, f_{D^*}  \, C^{(uc) VV/AV}_{\ell_1\ell_2}
\,,\quad 
g^{(V/A)}_{B^* \ell_1\ell_2} \, = \,  \frac{M_{B^*}^2}{\Lambda^2} 
\, f_{B^*}  \, C^{(db) VV/AV}_{\ell_1\ell_2}
\,,\nonumber
g^{(V/A)}_{B_s^* \ell_1\ell_2} \, = \,  \frac{M_{B_s^*}^2}{\Lambda^2} 
\, f_{B_s^*}  \, C^{(sb) VV/AV}_{\ell_1\ell_2}
\,,\nonumber\\ 
g^{(T)}_{\rho \ell_1\ell_2} &=&  
\frac{\hat{m} M_{\rho}}{\Lambda^{2}} \, f_{\rho} \, 
C^{(3) TT}_{\ell_1\ell_2} 
\,, \quad 
g^{(T)}_{\omega \ell_1\ell_2} \, = \, 
\frac{3 \hat{m} M_{\omega}}{\Lambda^{2}} \, f_{\omega} \, 
C^{(0) TT}_{\ell_1\ell_2} 
\,, \quad 
g^{(T)}_{\phi\ell_1\ell_2} \, = \,  
- \frac{3 m_s M_{\phi}}{\Lambda^{2}} \, f_{\phi} 
C^{(s) TT}_{\ell_1\ell_2} 
\,,\label{eq:matchingV} \\
g^{(T)}_{J/\psi\ell_1\ell_2} &=&
\frac{m_c M_{J/\psi}}{\Lambda^{2}} \, f_{J/\psi} 
C^{(c) TT}_{\ell_1\ell_2} 
\,, \quad 
g^{(T)}_{\Upsilon\ell_1\ell_2} \, = \,   
\frac{m_b M_{\Upsilon}}{\Lambda^{2}} \, f_{\Upsilon} 
C^{(b) TT}_{\ell_1\ell_2}
\,, \quad
g^{(T)}_{K^*\ell_1\ell_2} \, = \,   
\frac{(m_d+m_s) M_{K^*}}{2\Lambda^{2}} \, f_{K^*} 
C^{(ds) TT}_{\ell_1\ell_2}
\,, \nonumber\\
g^{(T)}_{D^*\ell_1\ell_2} &=&
\frac{(m_u+m_c) M_{D^*}}{2 \Lambda^{2}} \, f_{D^*} 
C^{(uc) TT}_{\ell_1\ell_2} 
\,, \quad 
g^{(T)}_{B^*\ell_1\ell_2} \, = \,   
\frac{(m_d+m_b) M_{B^*}}{2 \Lambda^{2}} \, f_{B^*} 
C^{(db) TT}_{\ell_1\ell_2}
\,,\nonumber\\
g^{(T)}_{B_s^*\ell_1\ell_2} \, & =&  \,   
\frac{(m_s+m_b) M_{\Upsilon}}{2 \Lambda^{2}} \, f_{B_s^*} 
C^{(sb) TT}_{\ell_1\ell_2}
\,. \nonumber
\end{eqnarray}

\noindent
\underline{Scalar mesons:}
\begin{eqnarray}
g^{(S/P)}_{a_0 \ell_1\ell_2} &=& 
\frac{M_{a_0}^2}{\Lambda^{2}} \, f_{a_0} \,
C^{(3) SS/PS}_{\ell_1\ell_2} 
\,,\quad 
g^{(S/P)}_{f_0 \ell_1\ell_2} \, = \,  
\frac{M_{f_0}^2}{\Lambda^{2}} \, f_{f_0} \,  
C^{(0) SS/PS}_{\ell_1\ell_2} 
\,,\quad 
g^{(S/P)}_{\chi_{c0} \ell_1\ell_2} \, = \,
\frac{M_{\chi_{c0}}^2}{\Lambda^{2}} \, f_{f_0} \,
C^{(c) SS/PS}_{\ell_1\ell_2}.\label{eq:matchingS}
\end{eqnarray}

\noindent
\underline{Pseudoscalar mesons:}
\begin{eqnarray}
g^{(P/S)}_{\pi \ell_1\ell_2} &=&  
\frac{M_\pi^2}{2 \hat{m} \Lambda^2} \, F_\pi \, 
C^{(3) PP/SP}_{\ell_1\ell_2} 
\,,\quad 
g^{(P/S)}_{\eta_c \ell_1\ell_2} \, = \,
\frac{M_{\eta_c}^2}{2 m_c \Lambda^2} \, F_{\eta_c} \, 
C^{(c) PP/SP}_{\ell_1\ell_2} 
\,,\nonumber\\ 
g^{(P/S)}_{\eta \ell_1\ell_2} &=& 
- \frac{M_\eta^2}{2 \hat{m} \Lambda^2} \, F_\pi \, 
\biggl( \sin\delta \,
C^{(0) PP/SP}_{\ell_1\ell_2} 
+ \frac{\hat{m}}{m_s} \, \cos\delta \, \sqrt{2} \, 
C^{(s) PP/SP}_{\ell_1\ell_2} 
\biggr)
\,,
\nonumber\\
g^{(P/S)}_{\eta' \ell_1\ell_2} && 
\frac{M_{\eta'}^2}{2 \hat{m} \Lambda^2} \, F_\pi \,
\biggl( \cos\delta \, 
C^{(0) PP/SP}_{\ell_1\ell_2} 
- \frac{\hat{m}}{m_s} \, \sin\delta \, \sqrt{2} \, 
C^{(s) PP/SP}_{\ell_1\ell_2} 
\biggr) 
\,,\nonumber\\ 
g^{(P/S)}_{K^0 \ell_1\ell_2} &=&  
\frac{M_{K^0}^2}{(m_d+m_s) \Lambda^2} \, F_{K^0} \, 
C^{(sd) PP/SP}_{\ell_1\ell_2}
\,,\quad 
g^{(P/S)}_{D^0 \ell_1\ell_2} \, = \,
\frac{M_{D^0}^2}{(m_u +m_c) \Lambda^2} \, F_{D^0} \, 
C^{(uc) PP/SP}_{\ell_1\ell_2}
\,,\nonumber\\ 
g^{(P/S)}_{B^0 \ell_1\ell_2} &=&  
\frac{M_{B^0}^2}{(m_u+m_b) \Lambda^2} \, F_{B^0} \, 
C^{(bd) PP/SP}_{\ell_1\ell_2}
\,,\quad 
g^{(P/S)}_{B^0_s \ell_1\ell_2} \, = \,
\frac{M_{B^0_s}^2}{(m_s +m_b) \Lambda^2} \, F_{B^0_s} \, 
C^{(bs) PP/SP}_{\ell_1\ell_2}
\,,\label{eq:matchingP}\\ 
g^{(A/V)}_{\pi \ell_1\ell_2} &=&  
- \frac{M_\pi}{\Lambda^2} \, F_\pi \, 
C^{(3) AA/VA}_{\ell_1\ell_2} 
\,,\quad 
g^{(A/V)}_{\eta_c \ell_1\ell_2} \, = \, 
- \frac{M_{\eta_c}}{\Lambda^2} \, F_{\eta_c} \, 
C^{(c) AA/VA}_{\ell_1\ell_2} \,,
\nonumber\\
g^{(A/V)}_{\eta \ell_1\ell_2} &=&    
 \frac{M_\eta}{\Lambda^2} \, F_\pi \, 
\biggl( \sin\delta \,
C^{(0) AA/VA}_{\ell_1\ell_2} 
\,+\, \cos\delta \, \sqrt{2} \, 
C^{(s) AA/VA}_{\ell_1\ell_2} 
\biggr)
\,,
\nonumber\\
g^{(A/V)}_{\eta' \ell_1\ell_2} &=&  
- \frac{M_{\eta'}}{\Lambda^2} \, F_\pi \, 
\biggl( \cos\delta \,
C^{(0) AA/VA}_{\ell_1\ell_2} 
\,-\, \sin\delta \, \sqrt{2} \, 
C^{(0) AA/VA}_{\ell_1\ell_2} 
\biggr) \,, 
\nonumber\\
g^{(A/V)}_{K^0 \ell_1\ell_2} &=&  
- \frac{M_{K^0}}{\Lambda^2} \, F_{K^0} \, 
C^{(ds) AA/VA}_{\ell_1\ell_2} 
\,,\quad 
g^{(A/V)}_{D^0 \ell_1\ell_2} \, = \, 
- \frac{M_{D^0}}{\Lambda^2} \, F_{D^0} \, 
C^{(cu) AA/VA}_{\ell_1\ell_2} \,,
\nonumber\\
g^{(A/V)}_{B^0 \ell_1\ell_2} &=&  
- \frac{M_{B^0}}{\Lambda^2} \, F_{B^0} \, 
C^{(bd) AA/VA}_{\ell_1\ell_2} 
\,,\quad 
g^{(A/V)}_{B^0_s \ell_1\ell_2} \, = \, 
- \frac{M_{B^0_s}}{\Lambda^2} \, F_{B^0_s} \, 
C^{(bs) AA/VA}_{\ell_1\ell_2} \,,
\nonumber
\en
where 
\eq 
C^{(0/3) \Gamma_i\Gamma_J}_{\ell_1\ell_2} = 
C^{(u) \Gamma_i\Gamma_J}_{\ell_1\ell_2} \pm 
C^{(d) \Gamma_i\Gamma_J}_{\ell_1\ell_2} 
\en 
is the strong isospin singlet ($C^{(0)}$) and 
triplet ($C^{(3)}$) combinations.
The dimensionless leptonic decay constant of the meson $M$ is denoted by $f_M$, and its mass by $M_M$.
Quark masses are represented by $m_q$, $\tilde m = m_u = m_d$ is the $u$ and $d$ quark in the isospin limit, and $\delta$ is related to the singlet-octet mixing angle.
More details and the values of the different constants can be found in Appendix \ref{app:constants}.

Now we substitute  the LFV meson-lepton effective couplings $g^{(M)}_{AB}$ in the decay rate formulas  (\ref{eq:rate-V-1})-(\ref{eq:Gamma-P}) by their expressions 
(\ref{eq:matchingV})-(\ref{eq:matchingP}) in terms of the WCs. In this way, we get  
for the LFV decay rates\\[3mm]
\underline{Vector mesons:}
\begin{align}
\label{eq:widthV}
\Gamma(V \to \ell_1^+ \ell_2^-) = & \,
\frac{|\vec{p}_\ell|}{6 \pi} \, 
  \biggl[
\biggl( 1 - \frac{m_-^2}{M_M^2} \biggr) \, 
\biggl( 1 + \frac{m_+^2}{2M_M^2} \biggr)  \,
\left(\alpha_{(q) \ell_1\ell_2}^{(V/A)}
C_{\ell_1\ell_2}^{(q)VV}\right)^2\nonumber\\
 & + 
\biggl( 1 - \frac{m_+^2}{M_M^2} \biggr) \, 
\biggl( 1 + \frac{m_-^2}{2M_M^2} \biggr)\,
\left(\alpha_{(q) \ell_1\ell_2}^{(V/A)}
C_{\ell_1\ell_2}^{(q)AV}\right)^2\nonumber\\
 & + 
2 \biggl( 1 - \frac{m_-^2}{M_M^2} \biggr) \, 
\biggl( 1 + \frac{2m_+^2}{M_M^2} \biggr) \,
\left(\alpha_{(q) \ell_1\ell_2}^{(T)}
C_{\ell_1\ell_2}^{(q)TT}\right)^2\nonumber\\
 & - \, 
6 \frac{m_+}{M_M} \, \biggl( 1 - \frac{m_-^2}{M_M^2} \biggr)
\left(\alpha_{(q) \ell_1\ell_2}^{(V/A)}\right)^2
\left(C_{\ell_1\ell_2}^{(q)VV} C_{\ell_1\ell_2}^{(q)AV} \right) 
\biggr],
\end{align}

\noindent
\underline{Scalar mesons:}
\small
\begin{align}\label{eq:widthS}
\Gamma(S \to \ell_1^+ \ell_2^-) = 
\frac{|\vec{p}_\ell|}{4 \pi} \, 
\left(\alpha_{(q) \ell_1\ell_2}^{(S/P)}\right)^2 
\biggl[
\biggl(1 - \frac{m_-^2}{M_M^2}\biggr)
\left( C_{\ell_1\ell_2}^{(q)SS}\right)^2
\,+\,  
\biggl(1 - \frac{m_+^2}{M_M^2}\biggr)
\left(C_{\ell_1\ell_2}^{(q)PS}\right)^2
\biggr],
\end{align}
\normalsize
\underline{Pseudoscalar mesons:}
\begin{align}\label{eq:widthP}
\Gamma(P\to \ell_1^+ \ell_2^-) = 
\frac{|\vec{p}_\ell|}{4 \pi} \, 
\biggl[
\biggl(1 - \frac{m_-^2}{M_M^2}\biggr)
\left(\alpha_{(q)\ell_1\ell_2}^{(P/S)} C_{\ell_1\ell_2}^{(q)PP}
  + \frac{m_+}{M_M} \alpha_{(q)\ell_1\ell_2}^{(A/V)} C_{\ell_1\ell_2}^{(q)AA} \right)^2\nonumber\\
\,+\, 
\biggl(1 - \frac{m_+^2}{M_M^2}\biggr)
\left(\alpha_{(q)\ell_1\ell_2}^{(P/S)} C_{\ell_1\ell_2}^{(q)SP} 
  + \frac{m_-}{M_M} \alpha_{(q)\ell_1\ell_2}^{(A/V)} C_{\ell_1\ell_2}^{(q)VA} \, 
\right)^2 \,  
\biggr],
\end{align}

\noindent
where the constants $\alpha_{(q_1 q_2)\ell_1\ell_2}^{(X)}$ are the factors multiplying the corresponding WCs in 
Eqs.~(\ref{eq:matchingV})-(\ref{eq:matchingP}), e.g., \mbox{$\alpha_{(3)\ell_1\ell_2}^{(V/A)} = \frac{m_\rho^2}{\Lambda^2}f_\rho$}. For other notations see Appendix \ref{App:M-rates}.

Using Eqs.~(\ref{eq:widthV})-(\ref{eq:widthP}) and the limits on the LFV meson decays from Table \ref{tab:existingbounds}, we extract limits on the WCs, assuming a single operator to contribute at a time.
In Tables \ref{tab:wcs1}-\ref{tab:wcs3nondiag} (second column)
we show these tree-level limits in terms of the effective parameters $\Lambda_{\ell_{1}\ell_{2}}^{(q_1 q_2)AB}$ defined as 
\begin{equation}\label{eq:wc-lambda-relation1}
\left|{C}^{(q_1 q_2)AB}_{\ell_{1}\ell_{2}}\right| \left(\frac{1\mbox{ GeV}}{\Lambda}\right)^2
  = 4 \pi \left( \frac{1\mbox{ GeV}}{\Lambda_{\ell_{1}\ell_{2}}^{(q_1 q_2)AB}} \right)^2,
\end{equation}
and characterizing the individual LFV scales of the operators in Eq.~(\ref{eq:l-q-operators-2}).
(We use the notation $C_{\ell_1\ell_2}^{(q_1 q_2)AB} \equiv C_{\ell_1\ell_2}^{(q)AB}$ when $q_1=q_2=q$.)
Below we examine the impact of the $\mbox{QED}\otimes\mbox{QCD}$ corrections on these tree-level limits.

\section{$QED\otimes QCD$ RGE improved Wilson Coefficients}
\label{sec:limitswithcorr}


Here we derive one-loop QED and QCD corrections to the LFV quark-lepton effective operators (\ref{eq:l-q-operators-2}) and find the renormalization group evolution (RGE) of the corresponding Wilson Coefficients.
We use the chiral basis of the LFV quark-lepton operators 
specified in Eqs.~(\ref{eq:Llq-12})-(\ref{eq:Chiral-Basis}) instead of the P-parity basis of Eq.~(\ref{eq:l-q-operators-2}). This basis is more convenient for our purposes and allows us to make a crosscheck with our previous similar calculations \cite{Gonzalez:2015ady,Gonzalez:2017mcg,Arbelaez:2016uto,Arbelaez:2016zlt}. Then we return to the basis of Eq.~(\ref{eq:l-q-operators-2}) using the relations (\ref{eq:op_basis_change-1})-(\ref{eq:op_basis_change-5}) .
\begin{figure}[tb]
\centering
\includegraphics[width=0.3\linewidth]{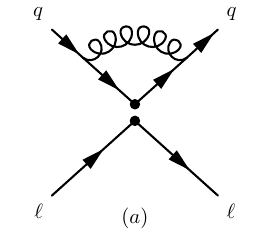}
\includegraphics[width=0.3\linewidth]{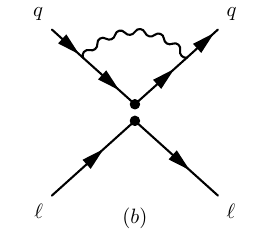}
\includegraphics[width=0.3\linewidth]{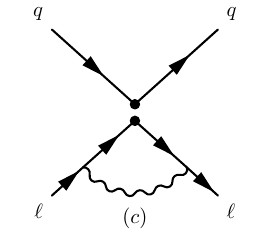}
\includegraphics[width=0.3\linewidth]{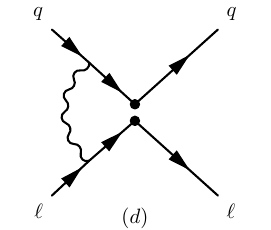}
\includegraphics[width=0.3\linewidth]{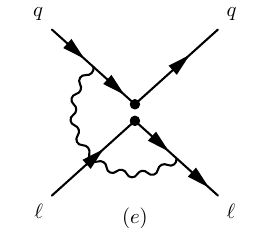}
\caption{One-loop QCD and QED corrections to the quark-lepton LFV effective operators. The contribution of the QCD-vertex correction in diagram ($a$) is shown in Eq.~(\ref{eq:Ma}), while the one-loop QED corrections in diagrams ($b$), ($c$), ($d$), and ($e$) are shown in Eqs.~(\ref{eq:Mb}), (\ref{eq:Mc}), (\ref{eq:Md}), and (\ref{eq:Me}), respectively.}
\label{fig:corrections}
\end{figure}

The one-loop corrections under consideration are shown in Fig.~\ref{fig:corrections}. 
Their $\overline{\rm MS}$ contributions to the operator matrix elements are \cite{Gonzalez:2015ady}
\begin{eqnarray}
\label{eq:Ma}
{\rm (a)} &\sim&    (\bar{\ell}\Gamma^i P_{X}\ell) \ \cdot \ (\bar q\gamma_{\alpha}\gamma_{\beta}\Gamma^j\gamma^{\beta}\gamma^{\alpha}P_{Y} q)\ \cdot \ C_F\frac{1}{4}\frac{\alpha_s}{4\pi}\left(\frac{1}{\epsilon}+\log\frac{\mu^2}{-p^2}\right),\\
\label{eq:Mb}
{\rm (b)} &\sim&    (\bar{\ell}\Gamma^i P_{X}\ell) \ \cdot \ (\bar q\gamma_{\alpha}\gamma_{\beta}\Gamma^j\gamma^{\beta}\gamma^{\alpha}P_{Y} q)\ \cdot \ Q_q^2\frac{1}{4}\frac{\alpha_{em}}{4\pi}\left(\frac{1}{\epsilon}+\log\frac{\mu^2}{-p^2}\right),\\
\label{eq:Mc}
{\rm (c)} &\sim&    (\bar{\ell}\gamma_{\alpha}\gamma_{\beta}\Gamma^i \gamma^{\beta}\gamma^{\alpha} P_{X}\ell) \ \cdot \ (\bar q\Gamma^jP_{Y} q)\ \cdot \ Q_\ell^2\frac{1}{4}\frac{\alpha_{em}}{4\pi}\left(\frac{1}{\epsilon}+\log\frac{\mu^2}{-p^2}\right),\\
\label{eq:Md}
{\rm (d)} &\sim&
    -(\bar{\ell}\Gamma^i\gamma_{\sigma}\gamma_{\alpha} P_{X} \ell) \ \cdot \ (\bar q\Gamma^j\gamma^{\sigma}\gamma^{\alpha} P_{Y}  q)\ \cdot \ Q_q Q_\ell\frac{1}{4}\frac{\alpha_{em}}{4\pi}\left(\frac{1}{\epsilon}+\log\frac{\mu^2}{-p^2}\right),
\\
\label{eq:Me}    
{\rm (e)} &\sim&
(\bar{\ell}\Gamma^i\gamma_{\sigma}\gamma_{\alpha} P_{X}  \ell) \ \cdot \ (\bar q\gamma^{\alpha}\gamma^{\sigma}\Gamma^j P_{Y}  q)\ \cdot \ Q_q Q_\ell \frac{1}{4}\frac{\alpha_{em}}{4\pi}\left(\frac{1}{\epsilon}+\log\frac{\mu^2}{-p^2}\right),
\end{eqnarray} 
where $\Gamma^i$ are the Lorentz structures corresponding to the operators in Eq.~(\ref{eq:Chiral-Basis}).
The general form of the corrected operator matrix elements is given by \cite{Gonzalez:2015ady}
\begin{eqnarray}
\langle\mathcal{O}_{i}\rangle^{(0)}&=& 
\left[\delta_{ij}+ 
\frac{\alpha_{s}}{4\pi}b_{ij}^s\left(\frac{1}{\epsilon}+\ln\left(\frac{\mu^2}{-p^2}\right)\right)
+
\frac{\alpha_{em}}{4\pi}b_{ij}^{em}\left(\frac{1}{\epsilon}+\ln\left(\frac{\mu^2}{-p^2}\right)\right)
\right]
\langle\mathcal{O}_{j}\rangle_{\rm tree},\label{eq:operatorME}
\end{eqnarray}
where $\langle\mathcal{O}_{i}\rangle_{\rm tree}$ are the operator matrix elements without $\qedqcd$ corrections. The $b_{ij}^s$ coefficients, characterizing the QCD corrections, form a  diagonal matrix 
\begin{eqnarray}
\hat b^{s,LL/RR} =
  \left(
  \begin{array}{ccc}
  4 C_F & 0 & 0\\
  0     & 0 & 0\\
  0 & 0 & C_F\end{array}
  \right),
\end{eqnarray}
\begin{eqnarray}
b_{11}^{s,LR/RL} = 4 C_F, \ \ b_{33}^{s,LR/RL} = C_F,
\end{eqnarray}
while the QED coefficients $b^{em}_{ij}$ have some non-diagonal matrix elements
\begin{eqnarray}
\hat b^{em,LL/RR} =
  \left(
  \begin{array}{ccc}
  4(Q_\ell^2+Q_q^2)& Q_q Q_\ell & 0\\
  48 Q_\ell Q_q & 0 & 0\\
  0 & 0 & Q_\ell^2 - 6 Q_\ell Q_q + Q_q^2
  \end{array}
  \right),
\end{eqnarray}
\begin{eqnarray}
b_{11}^{em,LR/RL} = 4(Q_\ell^2+Q_q^2),\ \ \ 
b_{33}^{em,LR/RL} = Q_\ell^2 + 6 Q_\ell Q_q + Q_q^2.
\end{eqnarray}

\subsection{Renormalization}
\noindent
The quark and lepton
field renormalization due to the $\qedqcd$ corrections is conventionally introduced by 
\begin{eqnarray}\label{eq:QuarkRen}
q^{(0)} = \mathcal{Z}_{q}^{1/2} Z_q^{1/2} q, \ \ \
\ell^{(0)} = Z_\ell^{1/2} \ell, \ \ \
\end{eqnarray}
where $q^{(0)}$, $\ell^{(0)}$ and $q$,$\ell$ are bare and renormalized quark fields, with the QCD and QED  renormalization functions given in the LO approximation, respectively by
\begin{eqnarray}
\mathcal{Z}_{q} = 1 - C_{F} \frac{\alpha_{s}}{4\pi} \, \frac{1}{\epsilon} + O(\alpha_{s}^{2}),\ \ \
Z_{q/\ell} = 1 - Q^2_{q/\ell} \frac{\alpha_{em}}{4\pi} \, \frac{1}{\epsilon} + O(\alpha_{em}^{2}).\ \ \
\end{eqnarray}
The
$1/\epsilon$-singularities in the operator matrix elements in Eq.~(\ref{eq:operatorME}) are removed by the quark and lepton field renormalization (\ref{eq:QuarkRen}), along with the renormalization of the operators, $\mathcal{O}_{i}^{(0)} =
Z_{ij}\mathcal{O}_{j}$.
Therefore, the operator matrix elements in Eq.~(\ref{eq:operatorME}) are renormalized in the following way
\begin{eqnarray}\label{eq:operatorMEren}
\langle\mathcal{O}_{i}\rangle^{(0)} = Z_{q}^{-1} Z_{\ell}^{-1} \mathcal{Z}_q^{-1} Z_{ij} \langle\mathcal{O}_{j}\rangle.
\end{eqnarray}
Requiring the cancelation of  the singularities in Eq.~(\ref{eq:operatorME}) one finds
\begin{eqnarray}\label{eq:Zij}
Z_{ij} = 
\delta_{ij} + 
\frac{\alpha_{s}}{4\pi} (b_{ij}^s -  C_{F} \delta_{ij}) 
\frac{1}{\epsilon}
+
\frac{\alpha_{em}}{4\pi} \left(b_{ij}^{em} -  (Q_q^2+Q_\ell^2) \delta_{ij} \right)  \frac{1}{\epsilon} +
O(\alpha_{s}^{2},\alpha_{em}\alpha_{s},\alpha_{em}^{2}) .
\end{eqnarray}

\subsection{Anomalous Dimensions}

The leading order expression for the anomalous dimension in the $\overline{\mbox{MS}}$-scheme is \cite{Buras:1998raa}   
\begin{equation}
\gamma_{ij}(\alpha_s,\alpha_{em})
	=  -2\alpha_s\frac{\partial \tilde Z^{em}_{ij}(\alpha_s,\alpha_{em})}{\partial \alpha_s}
	-2\alpha_{em}\frac{\partial \tilde Z^{s}_{ij}(\alpha_s,\alpha_{em})}{\partial \alpha_{em}},
\end{equation}
where $\tilde Z^{em,s}_{ij}$ are the factors multiplying $1/\epsilon$ in Eq.~(\ref{eq:Zij}).
Writing
\begin{eqnarray}
\gamma_{ij}(\alpha_S,\alpha_{em}) = \frac{\alpha_s}{4\pi}\gamma_{ij}^s + \frac{\alpha_{em}}{4\pi}\gamma_{ij}^{em},
\end{eqnarray}
we find
\begin{eqnarray}
\gamma_{ij}^s = -2(b_{ij}^s-C_F\delta_{ij}),\ \ \
\gamma_{ij}^{em} = -2\left(b_{ij}^{em}-(Q_q^2+Q_\ell^2)\delta_{ij}\right).
\end{eqnarray}
Using the previous equations, we find the one-loop anomalous dimensions
\begin{eqnarray}
&&\hat{\gamma}^{s,LL/RR}=-2\left(
\begin{array}{ccc}
3 C_F & 0 & 0 \\
0 & -C_F & 0\\
0 & 0 & 0
\end{array}
\right),
\end{eqnarray}
\begin{eqnarray}
\gamma^{s,LR/RL}_{11} = -6 C_F,\ \ \
\gamma^{s,LR/RL}_{33} = 0,
\end{eqnarray}
\begin{eqnarray}
&&\hat \gamma^{em,LL/RR}=\left(
\begin{array}{ccc}
-6 (Q_\ell^2 + Q_q^2) & -2 Q_\ell Q_q & 0 \\
-96\ Q_\ell Q_q & 2(Q_\ell^2 + Q_q^2) & 0\\
0 & 0 & 12\ Q_\ell Q_q
\end{array}
\right),
\end{eqnarray}
\begin{eqnarray}
\gamma_{11}^{em,LR/RL} = -6 (Q_\ell^2 + Q_q^2),\ \ \
\gamma_{33}^{em,LR/RL} = -12 Q_\ell Q_q.
\end{eqnarray}
The RG evolution of these operators has been already studied in the literature \cite{Davidson:2016edt,Jenkins:2017dyc}\footnote{We verified that our anomalous dimensions are in agreement with Ref.~\cite{Jenkins:2017dyc}, while showing a sign discrepancy with respect to Ref.~\cite{Davidson:2016edt} for the element $(\hat \gamma^{em,LL/RR})_{21}$}.
The solution to the RGE equations for the WCs
\begin{equation}
\frac{d \vec{C}(\mu)}{d\log \mu}
  = \hat\gamma^T \vec{C}(\mu)
\end{equation}
can be expressed as
\begin{equation}\label{eq:c-evolution}
\vec{C}(\mu) = \hat U(\mu,\Lambda)\, \vec{C}(\Lambda),
\end{equation}
where the evolution matrix $\hat U(\mu,\Lambda)$ has contributions from the QCD and QED corrections
\begin{equation}
\label{eq:RGE-U}
\hat U(\mu,\Lambda) = \hat U_s(\mu,\Lambda) + \hat U_{em}(\mu,\Lambda).
\end{equation}
The solution for the $\mu$-evolution matrices take the form \cite{Buchalla:1989we,Buras:1993dy,Aebischer:2017gaw} 
\begin{equation}
\hat U_s(\mu,\Lambda) = \hat V\left( \left[\frac{\alpha_s(\Lambda)}{\alpha_s(\mu)}\right]^{\frac{\hat \gamma^s}{2\beta_0}} \right)_D \hat V^{-1},
\end{equation}
where $\hat V$ is the matrix that diagonalizes the QCD anomalous dimension $\hat \gamma^s$
\begin{equation}
(\hat \gamma^s)_D = \hat V^{-1} \hat \gamma^{sT} \hat V,
\end{equation}
that in our case is just the identity matrix.
The QED contribution to the evolution operator is
\begin{equation}
\hat U_{em}(\mu,\Lambda) = - \frac{\alpha_{em}}{2\beta_0^s \alpha_s(\Lambda)} \hat V \hat K(\mu,\Lambda) \hat V^{-1},
\end{equation}
where de matrix elements of $\hat K(\mu,\Lambda)$ are given by
\begin{equation}
   K_{ij} (\mu,\Lambda) =
  ( \hat V^{-1} \hat \gamma^{emT} \hat V)_{ij} \times 
  \begin{cases}
    (\eta_s^{a_j+1}-\eta_s^{a_i})/(a_i-a_j-1)
   & \text{if } a_i-a_j \neq 1,\\
  \eta_s^{a_i}\ln \left( 1/\eta_s \right)
   & \text{if } a_i-a_j = 1,
  \end{cases}
\end{equation}
with
\begin{eqnarray}
\label{eq:RGE-coeff-eta-a}
\eta_s = \frac{\alpha_s(\Lambda)}{\alpha_s(\mu)}, \ \ \ a_i = \frac{(\hat \gamma^s)_{ii}}{2\beta_0^s}.
\end{eqnarray}
Using Eqs.~(\ref{eq:RGE-U})-(\ref{eq:RGE-coeff-eta-a}), we calculate the evolution matrix 
(\ref{eq:RGE-U}) in numerical form, first for the chiral basis (\ref{eq:Chiral-Basis}) and then, with the help of the relations (\ref{eq:op_basis_change-1})-(\ref{eq:op_basis_change-5}), for the basis of Eq.~(\ref{eq:l-q-operators-2}).
These numerical results are  listed in Appendix~\ref{app:evolution_dchirality} 
for $\mu=1\mbox{ GeV}$ and $\Lambda=1\mbox{ TeV}, 10\mbox{ TeV}$.

\section{RGE induced limits on the Wilson Coefficients}
\label{section:RGE_induced_limits}
The RGE evolution matrices $U_{ij}(\mu,\Lambda)$ calculated in the previous section allow us to obtain new limits on the WCs in Eq.~(\ref{eq:l-q-operators-2}) using as input the existing  tree-level limits in Refs.~\cite{Dib:2018rpy,Gutsche:2011bi,Nussinov:2000nm}\footnote{We updated some numerical values from Refs.~\cite{Dib:2018rpy,Gutsche:2011bi,Nussinov:2000nm}.} and the limits we  derived in section~\ref{sec:limitsnocorr}. 
We selected the most stringent bounds for each case and presented them in the second column of Tables \ref{tab:wcs1}-\ref{tab:wcs3nondiag}.

First, let us note that the effect of radiative corrections results in the replacements of the tree-level WCs, we denote by $C_{i}^{tree}$, in the formulas for the physical observables by their RGE-improved versions $C_{i}(\mu)$ given by (\ref{eq:c-evolution}). Therefore, the existing experimental upper limits derived at the tree level $C_{i}^{exp}$ are readily translated into limits on 
\begin{equation}
\label{eq:C-RGE-1}
C_i (\mu) = \sum_j U_{ij}(\mu,\Lambda)\, C_j (\Lambda) \leq C_{i}^{exp}.
\end{equation}
Nontrivial limits on $C_{j}(\Lambda)$ are obtained for those terms in the above expression corresponding to the non-diagonal matrix elements of  the RGE-evolution matrix $U_{ij}(\mu,\Lambda)$. 
Examining the numerical representation of this matrix 
shown in Appendix \ref{Sec:App-Non-Chiral}, we see that there are non-diagonal matrix elements corresponding to the  
mixing in the following subsets of the WCs $(SS,PP,TT)$, $(VV,AV)$ and $(AA,AV)$. 
We derive our limits on the basis of Eq.~(\ref{eq:C-RGE-1})  using the tree-level limits in the second column of  Tables \ref{tab:wcs1}-\ref{tab:wcs3nondiag}.
We conventionally apply an assumption about the presence of only a single term in the right-hand side of this equation at a time and derive simplified{\it ``on-axis''} limits on 
$C_{j}(\Lambda)$.
In the third and fourth columns of Tables \ref{tab:wcs1}-\ref{tab:wcs3nondiag},
we show our limits in terms of the individual LFV scales 
$\Lambda^{(q)AB}_{\ell_{1}\ell_{2}}$ defined in Eq.~(\ref{eq:wc-lambda-relation1}). 

\begin{table}
\centering
\begin{tabular}{c|c|c|c}
\ & Without  & With & With \\
\ $\Lambda_{\mu e}^{(q)}$ & $\mbox{QED}\otimes\mbox{QCD}$ & $\mbox{QED}\otimes\mbox{QCD}$ & $\mbox{QED}\otimes\mbox{QCD}$\\ 
\   & [TeV] & ($\Lambda=1\mbox{ TeV}$) [TeV] & ($\Lambda=10\mbox{ TeV}$) [TeV]\\ 
\hline\hline
$\Lambda_{\mu e}^{(0)VV/AV}$ & $5.1\times 10^3$ & $5.1\times 10^3$ & $5.1\times 10^3$\\
\hline
$\Lambda_{\mu e}^{(0)AA/VA}$ & $0.38$ & $4.6\times 10^{2}$ & $5.3\times 10^{2}$ \\
\hline
$\Lambda_{\mu e}^{(0)SS}$ & 1.8 & 23 &  27 \\
\hline
$\Lambda_{\mu e}^{(0)PS}$ & 1.8 & 2.7 & 2.8 \\
\hline
$\Lambda_{\mu e}^{(0)PP}$ & 5.4 & 23 & 27 \\
\hline
$\Lambda_{\mu e}^{(0)SP}$ & 5.4 & 7.8 & 8.3 \\
\hline
$\Lambda_{\mu e}^{(0)TT}$ & $5.8\times 10^{2}$ & $5.1\times 10^{2}$ & $5.0\times 10^{2}$ \\
\hline
\hline
$\Lambda_{\mu e}^{(3)VV/AV}$ & $5.4\times 10^{2}$ & $5.4\times 10^{2}$ & $5.4\times 10^{2}$ \\
\hline
$\Lambda_{\mu e}^{(3)AA/VA}$ & 2.2 & $8.0\times 10^{2}$ & $9.2\times 10^{2}$ \\
\hline
$\Lambda_{\mu e}^{(3)SS}$ & $0.45$ & 40 & 47 \\
\hline
$\Lambda_{\mu e}^{(3)PS}$ & $0.45$ & $0.65$ & $0.70$  \\
\hline
$\Lambda_{\mu e}^{(3)PP}$ & 8.0 & 40 & 47 \\
\hline
$\Lambda_{\mu e}^{(3)SP}$ & 7.9 & 11 & 12 \\
\hline
$\Lambda_{\mu e}^{(3)TT}$ & 61 & 54 & 53 \\
\hline
\hline
$\Lambda_{\mu e}^{(s)VV/AV}$ & $4.4\times 10^{2}$ & $4.4\times 10^{2}$ & $4.4\times 10^{2}$ \\
\hline
$\Lambda_{\mu e}^{(s)AA/VA}$ & 0.42 & 56 & 65 \\
\hline
$\Lambda_{\mu e}^{(s)SS}$ & no limit & 12 & 14 \\
\hline
$\Lambda_{\mu e}^{(s)PS}$ & no limit & $1.0\times 10^{-2}$ & $1.4\times 10^{-2}$ \\
\hline
$\Lambda_{\mu e}^{(s)PP}$ & 1.2 & 12 & 14 \\
\hline
$\Lambda_{\mu e}^{(s)SP}$ & 1.2 & 1.7 & 1.8 \\
\hline
$\Lambda_{\mu e}^{(s)TT}$ & $2.2\times 10^{2}$ & $1.9\times 10^{2}$ & $1.9\times 10^{2}$ \\
\hline
\hline
$\Lambda_{\mu e}^{(c)VV/AV}$ & 28 & 28 & 28 \\
\hline
$\Lambda_{\mu e}^{(c)AA/VA}$ & $3.0\times 10^{-2}$ & 5.1 & 5.8 \\
\hline
$\Lambda_{\mu e}^{(c)SS}$ & no limit & 1.7 & 2.0 \\
\hline
$\Lambda_{\mu e}^{(c)PS}$ & no limit & $3.0\times 10^{-3}$ & $4.2\times 10^{-3}$ \\
\hline
$\Lambda_{\mu e}^{(c)PP}$ & 0.17 & 1.7 & 2.0 \\
\hline
$\Lambda_{\mu e}^{(c)SP}$ & 0.17 & 0.25 & 0.27 \\
\hline
$\Lambda_{\mu e}^{(c)TT}$ & 21 & 19 & 18 \\
\hline
\hline
$\Lambda_{\mu e}^{(b)VV/AV}$ & 11 & 11 & 11 \\
\hline
$\Lambda_{\mu e}^{(b)AA/VA}$ & no limit & 1.4 & 1.7 \\
\hline
$\Lambda_{\mu e}^{(b)SS}$ & no limit & 0.52 & 0.60 \\
\hline
$\Lambda_{\mu e}^{(b)PS}$ & no limit & no limit & no limit \\
\hline
$\Lambda_{\mu e}^{(b)PP}$ & no limit & 0.52 & 0.60 \\
\hline
$\Lambda_{\mu e}^{(b)SP}$ & no limit & no limit & no limit \\
\hline
$\Lambda_{\mu e}^{(b)TT}$ & 9.2 & 8.1 & 8.0 \\
\hline
\hline
\end{tabular}
\caption{
Lower limits on the individual LFV scales $\Lambda_{\ell_1 \ell_2}^{(q)AB}$ ($\ell_1 \ell_2 = \mu e$) associated to the Wilson coefficient $C_{\ell_1 \ell_2}^{(q)AB}$.
For the definition of the LFV scale, see Eq.~(\ref{eq:wc-lambda-relation1}).
The limits without corrections are obtained from the limits on the LFV decays, $M\to \ell_1^\pm \ell_2^\mp$ ($M=\pi^0,\eta,\eta',\eta_c,f_0,a_0,\chi_{c0},\rho,\omega,\phi,J/\psi,\Upsilon$), listed in Table \ref{tab:existingbounds}.
The improvement of the limits in the presence of $\mbox{QED}\otimes\mbox{QCD}$ corrections is due to operator mixing.
The mixing among different operators (RGE evolution matrices) can be found in Appendix \ref{Sec:App-Non-Chiral}.
}\label{tab:wcs1}
\end{table}

\begin{table}
\small
\centering
\begin{tabular}{c|c|c|c}
\ & Without  & With & With \\
\ $\Lambda_{\tau e}^{(q)}$ & $\mbox{QED}\otimes\mbox{QCD}$ & $\mbox{QED}\otimes\mbox{QCD}$ & $\mbox{QED}\otimes\mbox{QCD}$\\ 
\   & [TeV] & ($\Lambda=1\mbox{ TeV}$) [TeV] & ($\Lambda=10\mbox{ TeV}$) [TeV]\\ 
\hline\hline
$\Lambda_{\tau e}^{(c)VV/AV}$ & 1.2 & 1.2 & 1.2 \\
\hline
$\Lambda_{\tau e}^{(c)AA/VA}$ & no limit & 0.22 & 0.25\\
\hline
$\Lambda_{\tau e}^{(c)SS}$ & no limit & $8.3\times 10^{-2}$ & $9.7\times 10^{-2}$\\
\hline
$\Lambda_{\tau e}^{(c)PP}$ & no limit & $8.3\times 10^{-2}$ & $9.7\times 10^{-2}$\\
\hline
$\Lambda_{\tau e}^{(c)TT}$ & 1.0 & 0.92 & 0.90\\
\hline
\hline
$\Lambda_{\tau e}^{(b)VV/AV}$ & $0.59$ & $0.59$ & $0.59$\\
\hline
$\Lambda_{\tau e}^{(b)AA/VA}$ & no limit & $7.4\times 10^{-2}$ & $8.5\times 10^{-2}$\\
\hline
$\Lambda_{\tau e}^{(b)SS}$ & no limit & $2.6\times 10^{-2}$ & $3.1\times 10^{-2}$\\
\hline
$\Lambda_{\tau e}^{(b)PP}$ & no limit & $2.6\times 10^{-2}$ & $3.1\times 10^{-2}$\\
\hline
$\Lambda_{\tau e}^{(b)TT}$ & $0.47$ & $0.41$ & $0.41$\\
\hline
\hline
\end{tabular}
\caption{
Lower limits on the individual LFV scales $\Lambda_{\ell_1 \ell_2}^{(q)AB}$ ($\ell_1 \ell_2 = \tau e$) associated to the wilson coefficient $C_{\ell_1 \ell_2}^{(q)AB}$.
See the caption of Table \ref{tab:wcs1} for further details.
}
\label{tab:wcs2}
\end{table}

\begin{table}
\centering
\begin{tabular}{c|c|c|c}
\ & Without  & With & With \\
\ $\Lambda_{\tau \mu}^{(q)}$ & $\mbox{QED}\otimes\mbox{QCD}$ & $\mbox{QED}\otimes\mbox{QCD}$ & $\mbox{QED}\otimes\mbox{QCD}$\\ 
\   & [TeV] & ($\Lambda=1\mbox{ TeV}$) [TeV] & ($\Lambda=10\mbox{ TeV}$) [TeV]\\ 
\hline\hline
$\Lambda_{\tau \mu}^{(c)VV/AV}$ & $1.3$ & $1.3$ & $1.3$\\
\hline
$\Lambda_{\tau \mu}^{(c)AA/VA}$ & no limit & $0.24$ & 0.$28$\\
\hline
$\Lambda_{\tau \mu}^{(c)SS}$ & no limit & $9.4\times 10^{-2}$ & $0.11$\\
\hline
$\Lambda_{\tau \mu}^{(c)PP}$ & no limit & $9.4\times 10^{-2}$ & $0.11$\\
\hline
$\Lambda_{\tau \mu}^{(c)TT}$ & $1.1$ & $1.0$ & $1.0$\\
\hline
\hline
$\Lambda_{\tau \mu}^{(b)VV/AV}$ & $0.64$ & $0.64$ & $0.64$\\
\hline
$\Lambda_{\tau \mu}^{(b)AA/VA}$ & no limit & $8.2\times 10^{-2}$ & $9.4\times 10^{-2}$\\
\hline
$\Lambda_{\tau \mu}^{(b)SS}$ & no limit & $2.9\times 10^{-2}$ & $3.4\times 10^{-2}$\\
\hline
$\Lambda_{\tau \mu}^{(b)PP}$ & no limit & $2.9\times 10^{-2}$ & $3.4\times 10^{-2}$\\
\hline
$\Lambda_{\tau \mu}^{(b)TT}$ & $0.52$ & $0.46$ & $0.45$\\
\hline
\hline
\end{tabular}
\caption{
Lower limits on the individual LFV scales $\Lambda_{\ell_1 \ell_2}^{(q)AB}$ ($\ell_1 \ell_2 = \tau \mu$) associated to the wilson coefficient $C_{\ell_1 \ell_2}^{(q)AB}$.
See the caption of Table \ref{tab:wcs1} for further details.
}
\label{tab:wcs3} 
\end{table}

\begin{table}
\centering
\begin{tabular}{c|c|c|c}
\ & Without  & With & With \\
\ $\Lambda_{\mu e}^{(q_1 q_2)}$ & $\mbox{QED}\otimes\mbox{QCD}$ & $\mbox{QED}\otimes\mbox{QCD}$ & $\mbox{QED}\otimes\mbox{QCD}$\\ 
\   & [TeV] & ($\Lambda=1\mbox{ TeV}$) [TeV] & ($\Lambda=10\mbox{ TeV}$) [TeV]\\ 
\hline
\hline
$\Lambda_{\mu e}^{(ds)VV/AV}$ & $34$ & $2.0\times 10^2$ & $2.3\times 10^2$ \\
\hline
$\Lambda_{\mu e}^{(ds)AA/VA}$ & $1.5\times 10^3$ & $1.5\times 10^3$ & $1.5\times 10^3$ \\
\hline
$\Lambda_{\mu e}^{(ds)SS/PS}$ & no limit & $49$ & $68$ \\
\hline
$\Lambda_{\mu e}^{(ds)PP/SP}$ & $5.7\times 10^3$  & $8.2\times 10^3$ & $8.8\times 10^3$ \\
\hline
$\Lambda_{\mu e}^{(ds)TT}$ & $13$ & $1.7\times 10^3$ & $2.1\times 10^3$ \\
\hline
\hline
$\Lambda_{\mu e}^{(cu)VV/AV}$ & $24$ & $24$ & $23$ \\
\hline
$\Lambda_{\mu e}^{(cu)AA/VA}$ & $22$ & $22$ & $22$ \\
\hline
$\Lambda_{\mu e}^{(cu)SS/PS}$ & no limit & $1.9$ & $2.7$ \\
\hline
$\Lambda_{\mu e}^{(cu)PP/SP}$ & $1.1\times 10^2$ & $1.6\times 10^2$  & $1.7\times 10^2$ \\
\hline
$\Lambda_{\mu e}^{(cu)TT}$ & $16$ & $49$ & $59$ \\
\hline
\hline
$\Lambda_{\mu e}^{(db)VV/AV}$ & $5.3$ & $9.3$ & $10$ \\
\hline
$\Lambda_{\mu e}^{(db)AA/VA}$ & $73$ & $73$ & $73$ \\
\hline
$\Lambda_{\mu e}^{(db)SS/PS}$ & no limit & $5.0$ & $6.9$ \\
\hline
$\Lambda_{\mu e}^{(db)PP/SP}$ & $5.8\times 10^2$ & $8.5\times 10^2$ & $9.0\times 10^2$ \\
\hline
$\Lambda_{\mu e}^{(db)TT}$ & $3.9$ & $1.8\times 10^2$  & $2.1\times 10^2$ \\
\hline
\hline
$\Lambda_{\mu e}^{(sb)VV/AV}$ & $6.4$ & $6.7$ & $7.8$ \\
\hline
$\Lambda_{\mu e}^{(sb)AA/VA}$ & $53$ & $53$ & $53$ \\
\hline
$\Lambda_{\mu e}^{(sb)SS/PS}$ & no limit & $3.6$ & $5.0$ \\
\hline
$\Lambda_{\mu e}^{(sb)PP/SP}$ & $4.2\times 10^{2}$ & $6.1\times 10^2$ & $6.5\times 10^2$ \\
\hline
$\Lambda_{\mu e}^{(sb)TT}$ & $4.8$ & $1.3\times 10^2$ & $1.5\times 10^2$\\
\hline\hline
\end{tabular}
\caption{
Lower limits on the individual LFV scales $\Lambda_{\ell_1 \ell_2}^{(q_1 q_2)AB}$ ($q_1\neq q_2$, $\ell_1 \ell_2 = \mu e$) associated to the wilson coefficient $C_{\ell_1 \ell_2}^{(q_1 q_2)AB}$.
For the definition of the LFV scale, see Eq.~(\ref{eq:wc-lambda-relation1}).
The limits without corrections are obtained from the limits on the LFV decays, $M\to \ell_1^\pm \ell_2^\mp$ ($M=K_L^0,D^0,B^0,B^0_s,K^*,D^*,B^*.B^*_s$), listed in Table \ref{tab:existingbounds}.
The improvement of the limits in the presence of $\mbox{QED}\otimes\mbox{QCD}$ corrections is due to operator mixing.
The mixing among different operators (RGE evolution matrices) can be found in Appendix \ref{Sec:App-Non-Chiral}.
}\label{tab:wcs1nondiag}
\end{table}

\begin{table}
\centering
\begin{tabular}{c|c|c|c}
\ & Without  & With & With \\
\ $\Lambda_{\tau e}^{(q_1 q_2)}$ & $\mbox{QED}\otimes\mbox{QCD}$ & $\mbox{QED}\otimes\mbox{QCD}$ & $\mbox{QED}\otimes\mbox{QCD}$\\ 
\   & [TeV] & ($\Lambda=1\mbox{ TeV}$) [TeV] & ($\Lambda=10\mbox{ TeV}$) [TeV]\\ 
\hline
\hline
$\Lambda_{\tau e}^{(cu)VV/AV}$ & $0.61$ & $0.61$ & $0.61$ \\
\hline
$\Lambda_{\tau e}^{(cu)AA/VA}$ & no limit & $0.11$ & $0.12$ \\
\hline
$\Lambda_{\tau e}^{(cu)SS}$ & no limit & $3.8\times 10^{-2}$ & $4.5\times 10^{-2}$ \\
\hline
$\Lambda_{\tau e}^{(cu)PP}$ & no limit & $3.8\times 10^{-2}$ & $4.5\times 10^{-2}$ \\
\hline
$\Lambda_{\tau e}^{(cu)TT}$ & $0.48$ & $0.42$ & $0.41$ \\
\hline
\hline
$\Lambda_{\tau e}^{(db)VV/AV}$ & $0.25$ & $2.8$ & $3.2$ \\
\hline
$\Lambda_{\tau e}^{(db)AA/VA}$ & $22$ & $22$ & $22$ \\
\hline
$\Lambda_{\tau e}^{(db)SS/PS}$ & no limit & $0.36$ & $0.50$ \\
\hline
$\Lambda_{\tau e}^{(db)PP/SP}$ & $42$ & $61$ & $65$ \\
\hline
$\Lambda_{\tau e}^{(db)TT}$ & $0.20$ & $13$  & $15$ \\
\hline
\hline
$\Lambda_{\tau e}^{(sb)VV/AV}$ & $0.31$ & $2.8$ & $3.2$ \\
\hline
$\Lambda_{\tau e}^{(sb)AA/VA}$ & $22$ & $22$ & $22$ \\
\hline
$\Lambda_{\tau e}^{(sb)SS/PS}$ & no limit & $0.36$ & $0.50$ \\
\hline
$\Lambda_{\tau e}^{(sb)PP/SP}$ & $42$ & $61$ & $65$ \\
\hline
$\Lambda_{\tau e}^{(sb)TT}$ & $0.24$ & $13$ & $15$\\
\hline\hline
\end{tabular}
\caption{
Lower limits on the individual LFV scales $\Lambda_{\ell_1 \ell_2}^{(q_1 q_2)AB}$ ($q_1\neq q_2$, $\ell_1 \ell_2 = \tau e$) associated to the wilson coefficient $C_{\ell_1 \ell_2}^{(q_1 q_2)AB}$.
See the caption of Table \ref{tab:wcs1nondiag} for further details.
}\label{tab:wcs2nondiag}
\end{table}

\begin{table}
\centering
\begin{tabular}{c|c|c|c}
\ & Without  & With & With \\
\ $\Lambda_{\tau \mu}^{(q_1 q_2)}$ & $\mbox{QED}\otimes\mbox{QCD}$ & $\mbox{QED}\otimes\mbox{QCD}$ & $\mbox{QED}\otimes\mbox{QCD}$\\ 
\   & [TeV] & ($\Lambda=1\mbox{ TeV}$) [TeV] & ($\Lambda=10\mbox{ TeV}$) [TeV]\\ 
\hline
\hline
$\Lambda_{\tau \mu}^{(cu)VV}$ & $0.73$ & $0.73$ & $0.73$ \\
\hline
$\Lambda_{\tau \mu}^{(cu)AV}$ & $0.56$ & $0.56$ & $0.56$ \\
\hline
$\Lambda_{\tau \mu}^{(cu)AA}$ & no limit & $0.13$ & $0.15$ \\
\hline
$\Lambda_{\tau \mu}^{(cu)VA}$ & no limit & $0.10$ & $0.11$ \\
\hline
$\Lambda_{\tau \mu}^{(cu)SS}$ & no limit & $4.6\times 10^{-2}$ & $5.3\times 10^{-2}$ \\
\hline
$\Lambda_{\tau \mu}^{(cu)PP}$ & no limit & $4.6\times 10^{-2}$ & $5.3\times 10^{-2}$ \\
\hline
$\Lambda_{\tau \mu}^{(cu)TT}$ & $0.57$ & $0.51$ & $0.49$ \\
\hline
\hline
$\Lambda_{\tau \mu}^{(db)VV}$ & $0.28$ & $3.2$ & $3.9$ \\
\hline
$\Lambda_{\tau \mu}^{(db)AV}$ & $0.28$ & $3.4$ & $3.7$ \\
\hline
$\Lambda_{\tau \mu}^{(db)AA}$ & $27$ & $27$ & $27$ \\
\hline
$\Lambda_{\tau \mu}^{(db)VA}$ & $25$ & $25$ & $25$ \\
\hline
$\Lambda_{\tau \mu}^{(db)SS/PS}$ & no limit & $0.43$ & $0.60$ \\
\hline
$\Lambda_{\tau \mu}^{(db)PP/SP}$ & $50$ & $73$ & $78$ \\
\hline
$\Lambda_{\tau \mu}^{(db)TT}$ & $0.22$ & $15$  & $18$ \\
\hline
\hline
$\Lambda_{\tau \mu}^{(sb)VV/AV}$ & $0.34$ & $0.34$ & $0.34$ \\
\hline
$\Lambda_{\tau \mu}^{(sb)AA/VA}$ & no limit & $4.3\times 10^{-2}$ & $5.0\times 10^{-2}$ \\
\hline
$\Lambda_{\tau \mu}^{(sb)SS}$ & no limit & $1.5\times 10^{-2}$ & $1.7\times 10^{-2}$ \\
\hline
$\Lambda_{\tau \mu}^{(sb)PP}$ & no limit & $1.5\times 10^{-2}$ & $1.7\times 10^{-2}$ \\
\hline
$\Lambda_{\tau \mu}^{(sb)TT}$ & $0.27$ & $0.24$ & $0.23$\\
\hline\hline
\end{tabular}
\caption{
Lower limits on the individual LFV scales $\Lambda_{\ell_1 \ell_2}^{(q_1 q_2)AB}$ ($q_1\neq q_2$, $\ell_1 \ell_2 = \tau \mu$) associated to the Wilson Coefficient $C_{\ell_1 \ell_2}^{(q_1 q_2)AB}$.
See the caption of Table \ref{tab:wcs1nondiag} for further details.
}\label{tab:wcs3nondiag}
\end{table}

\section{Summary and Conclusions}\label{sec:conclusions}

Applying unitary-based arguments, we derived new upper limits on the leptonic decays of flavored mesons from the experimental limits on three-body LFV leptonic decays of $\mu$ and $\tau$. We also updated the existing phenomenological limits of  the unflavored mesons. These limits are shown in Table~\ref{tab:existingbounds}.

We studied the effective lepton-quark dimension-6 LFV operators (\ref{eq:l-q-operators-2}) and their hadronization via embedding the quark currents into (axial-)vector, (pseudo-)scalar mesons as shown in Eq.~(\ref{eq:L-meson-lepton}).
The embedding was realized by matching the quark-lepton (\ref{eq:l-q-operators-2}) and meson-lepton Lagrangians (\ref{eq:L-meson-lepton}) on mass-shell. This allowed us to express leptonic decay rates of mesons in terms of  the WCs of the quark-lepton Lagrangian and extract upper limits on them from the previously derived limits on the meson decay rates. Some of these extracted limits are updated versions of the ones existing in the literature. New limits, we derived in the present paper, are those for the WCs $C_{\ell_1\ell_2}^{(q_1 q_2)AB}$ ($q_1\neq q_2$) corresponding to the quark-flavor non-diagonal operators. These new tree-level limits were derived from the LFV decays of the vector flavored mesons $K^*,D^*,B^*,B_s^*$ and pseudoscalar flavored mesons  $K^0,D^0,B^0,B^0_s$.

Then we analyzed the effects of QED and QCD corrections to the effective quark-lepton LFV Lagrangian (\ref{eq:l-q-operators-2}).
We derived all the RGE-evolution matrices relevant for the considered operators.
These matrices are shown Appendix \ref{Sec:App-Non-Chiral}.

We found that QED corrections are particularly important due to operator mixing.
The elements of the operator basis, which mix with each other, correspond to the subsets $(C^{(q_1 q_2)SS},C^{(q_1 q_2)PP},C^{(q_1 q_2)TT})$, $(C^{(q_1 q_2)AA},C^{(q_1 q_2)VV})$, and $(C^{(q_1 q_2)AV},C^{(q_1 q_2)VA})$.
There is an additional mixing between the singlet and triplet isospin combinations, \mbox{$C^{(0/3)\Gamma_i\Gamma_j} = C^{(u)\Gamma_i\Gamma_j} \pm C^{(d)\Gamma_i\Gamma_j}$}, because operators with u-type and d-type quarks evolve differently under QED due to their different charges.

The mixing within the operator subsets $(C^{(q_1 q_2)AA},C^{(q_1 q_2)VV})$ and $(C^{(q_1 q_2)AV},C^{(q_1 q_2)VA})$ is responsible for the significant improvement (up to three orders of magnitude) of the $AA/VA$ limits.
This happens because the operator mixing makes vector meson decays sensitive not only to
the $VV/AV$ WCs, as at tree level, but also to the $AA/VA$. As a result, the strong limits on vector meson decays provide the above-mentioned improvement of the $AA/VA$ limits.
The limits on $ SS / PS $ WCs have also been tightened by up to two orders of magnitude.
In addition, these bounds, which are almost the same among themselves without taking into account the amendments of $\qedqcd$, after the introduction of corrections are split into different constraints for $ SS $ and $ PS $.
This is because the $SS$ couplings mix with the tensor $TT$, while $PS$ couplings do not.
Finally, the limits on $TT$ operators are slightly weakened, mainly due to QCD effects.

Our constraints can be useful for further studies of phenomenological implications of the effective LFV Lagrangian (\ref{eq:l-q-operators-2}) and its ultraviolet completions.

\bigskip

\centerline{\bf Acknowledgements}

\medskip

This work was supported by ANID-Chile Fondecyt 1190845, ANID-Chile Fondecyt Iniciaci\'on 11180873, ANID PIA/APOYO AFB180002 and Milenio-ANID-ICN2019\_044 as well as by ANID REC Convocatoria Nacional Subvenci\'on a Instalaci\'on en la Academia Convocatoria A\~no 2020, PAI77200092.

\newpage

\setcounter{section}{0}
\section*{Appendices}
\def\theequation{\Alph{section}.\arabic{equation}}
\begin{appendix}
\setcounter{equation}{0}

\section{Leptonic Decay Rates of Mesons} 
\label{App:M-rates}

Here we present analytical expressions for the LFV decay rates of mesons decaying into a leptonic pair governed by the effective 
Lagrangian~(\ref{eq:L-meson-lepton})

\vspace{3mm}
\noindent
Vector mesons:
\eq
\label{eq:rate-V-1} 
\Gamma(V \to \ell_1^+ \ell_2^-) &=& 
\frac{|\vec{p}_\ell|}{6 \pi} \, 
  \biggl[
\biggl(g_{V \ell_1\ell_2}^{(V)}\biggr)^2 \ 
\biggl( 1 - \frac{m_-^2}{M_V^2} \biggr) \, 
\biggl( 1 + \frac{m_+^2}{2M_V^2} \biggr)  \, + \, 
\biggl(g_{V \ell_1\ell_2}^{(A)}\biggr)^2 \ 
\biggl( 1 - \frac{m_+^2}{M_V^2} \biggr) \, 
\biggl( 1 + \frac{m_-^2}{2M_V^2} \biggr)  \nonumber\\
&+& 
2 \biggl(g_{V \ell_1\ell_2}^{(T)}\biggr)^2 \ 
\biggl( 1 - \frac{m_-^2}{M_V^2} \biggr) \, 
\biggl( 1 + \frac{2m_+^2}{M_V^2} \biggr)  \, - \, 
6 g_{V \ell_1\ell_2}^{(V)} \, g_{V \ell_1\ell_2}^{(A)} \, 
\frac{m_+}{M_V} \, \biggl( 1 - \frac{m_-^2}{M_V^2} \biggr) 
\biggr].
\en 

\noindent
Scalar mesons:
\eq
\Gamma(S \to \ell_1^+ \ell_2^-) = 
\frac{|\vec{p}_\ell|}{4 \pi} \, 
\biggl[\biggl(g_{S\ell_1\ell_2}^{(S)}\biggr)^2 
\, \biggl(1 - \frac{m_-^2}{M_S^2}\biggr)
\,+\, 
\biggl(g_{S\ell_1\ell_2}^{(P)}\biggr)^2 
\, \biggl(1 - \frac{m_+^2}{M_S^2}\biggr)
\biggr] .
\en 

\noindent
Pseudoscalar mesons:
\eq
\label{eq:Gamma-P}
\Gamma(P \to \ell_1^+ \ell_2^-) = 
\frac{|\vec{p}_\ell|}{4 \pi} \, 
\biggl[\biggl(g_{P\ell_1\ell_2}^{(P)} + g_{P\ell_1\ell_2}^{(A)} \, 
\frac{m_+}{M_P}\biggr)^2 \, \biggl(1 - \frac{m_-^2}{M_P^2}\biggr) 
\,+\, 
\biggl(g_{P\ell_1\ell_2}^{(S)} + g_{P\ell_1\ell_2}^{(V)} \, 
\frac{m_-}{M_P}\biggr)^2 \, \biggl(1 - \frac{m_+^2}{M_P^2}\biggr) 
\biggr],
\en 
with $m_\pm = m_{\ell_1} \pm m_{\ell_2}$ and
$|\vec{p}_\ell| = \lambda^{1/2}(M_M^2,m_{\ell_1}^2,m_{\ell_2}^2)/(2 M_M)$ is 
the magnitude of the momentum of the leptons in the rest 
frame of the decaying meson, where $\lambda(x,y,z) = x^2 + y^2 + z^2 - 2 xy - 2 yz - 2 xz$.

\section{Chiral Operator Basis}
\label{Appendix:Chiral-Basis}
The LFV quark-lepton effective Lagrangian (\ref{eq:l-q-operators-2}) 
\begin{eqnarray}\label{eq:Llq-12} 
{\cal L}_{\ell q} = \frac{1}{\Lambda^2} \sum_{i, XY} C_{i,\,\ell_1\ell_2}^{(q_1 q_2)XY}(\mu)\cdot \mathcal{O}^{(q_1 q_2)XY}_{i,\,\ell_1\ell_2}(\mu) + \mbox{H.c.},
\end{eqnarray}
is frequently expressed in the literature 
\cite{Gonzalez:2015ady,Buras:1998raa,Buchalla:1989we,Buras:1993dy,Aebischer:2017gaw}
in the basis of chiral operators
\begin{eqnarray}
\label{eq:Chiral-Basis} 
\mathcal{O}^{(q_1 q_2)XY}_{1,\,\ell_1\ell_2} & = & 4 ({\bar \ell_1}P_{X}\ell_2) ({\bar q_1}P_{Y}q_2),\nonumber\\
\mathcal{O}^{(q_1 q_2)XX}_{2,\,\ell_1\ell_2}&=& 4 ({\bar \ell_1}\sigma^{\mu\nu}P_{X}\ell_2) ({\bar q_1}\sigma_{\mu\nu}P_{X}q_2),\\ \label{eq:opbasis}
\mathcal{O}^{(q_1 q_2)XY}_{3,\,\ell_1\ell_2}&=& 4 ({\bar \ell_1}\gamma^{\mu}P_{X}\ell_2) 
                        ({\bar q_1}\gamma_{\mu}P_{Y}q_2),\nonumber
\end{eqnarray}
with  $X,Y = L,R$ and $P_{L/R} = (1\mp \gamma_{5})/2$. Our RGE calculations in Sec.~\ref{sec:limitswithcorr} are carried out in this basis. 
The WCs ${C}^{(q_1 q_2)XY}_{i\,,\ell_1\ell_2}$ in Eq.~(\ref{eq:Llq-12})
are related with the WCs $C^{(q_1 q_2)AB}_{\ell_{1}\ell_{2}}$ in Eq.~(\ref{eq:l-q-operators-2})  as follows
\begin{eqnarray}\label{eq:op_basis_change-1}
{C}^{(q_1 q_2)SS/VV}_{\ell_1\ell_2} & = & C_{1/3,\,\ell_1\ell_2}^{(q_1 q_2)LL} + C_{1/3,\,\ell_1\ell_2}^{(q_1 q_2)LR} + C_{1/3,\,\ell_1\ell_2}^{(q_1 q_2)RL} + C_{1/3,\,\ell_1\ell_2}^{(q_1 q_2)RR},\\
\label{eq:op_basis_change-2}
{C}^{(q_1 q_2)PS/AV}_{\ell_1\ell_2} & = &-C_{1/3,\,\ell_1\ell_2}^{(q_1 q_2)LL} - C_{1/3,\,\ell_1\ell_2}^{(q_1 q_2)LR} + C_{1/3,\,\ell_1\ell_2}^{(q_1 q_2)RL} + C_{1/3,\,\ell_1\ell_2}^{(q_1 q_2)RR},\\
\label{eq:op_basis_change-3}
{C}^{(q_1 q_2)SP/VA}_{\ell_1\ell_2} & = &-C_{1/3,\,\ell_1\ell_2}^{(q_1 q_2)LL} + C_{1/3,\,\ell_1\ell_2}^{(q_1 q_2)LR} - C_{1/3,\,\ell_1\ell_2}^{(q_1 q_2)RL} + C_{1/3,\,\ell_1\ell_2}^{(q_1 q_2)RR},\\
\label{eq:op_basis_change-4}
{C}^{(q_1 q_2)PP/AA}_{\ell_1\ell_2} & = & C_{1/3,\,\ell_1\ell_2}^{(q_1 q_2)LL} - C_{1/3,\,\ell_1\ell_2}^{(q_1 q_2)LR} - C_{1/3,\,\ell_1\ell_2}^{(q_1 q_2)RL} + C_{1/3,\,\ell_1\ell_2}^{(q_1 q_2)RR},\\
\label{eq:op_basis_change-5}     
{C}^{(q_1 q_2)TT}_{\ell_1\ell_2}\hspace{6mm} & = & 4\,C_{2,\,\ell_1\ell_2}^{(q_1 q_2)LL} = 4\,C_{2,\,\ell_1\ell_2}^{(q_1 q_2)RR}.
\end{eqnarray}
We use these relations to find the RGE evolution matrices for the non-chiral operator basis 
(\ref{eq:l-q-operators-2}) starting from the corresponding numerical evolution matrices for the chiral basis.

\section{RGE-evolution Matrices in the Chiral Basis}
\label{app:evolution_dchirality}

Here we give the RGE evolution matrices, $U(\mu,\Lambda)$, in the basis of Eq.~(\ref{eq:Chiral-Basis}) for 
$\mu=1\mbox{ GeV}$ and two values of the cutoff scale $\Lambda=1\mbox{ TeV},10\mbox{ TeV}$.

\subsection{$\Lambda=1\mbox{ TeV}$}

\begin{align}
U^{(U)LL/RR} = \left(
\begin{array}{ccc}
 2.104 & -0.377 & 0. \\
 -0.006 & 0.781 & 0. \\
 0. & 0. & 1.032 \\
\end{array}
\right),\ \ \
U_1^{(U)LR/RL} = 2.103,\ \ \
U_3^{(U)LR/RL} = 0.968,
\label{eq:ULLRR}
\end{align}
\begin{align}
U^{(D)LL/RR} = \left(
\begin{array}{ccc}
 2.087 & 0.188 & 0. \\
 0.003 & 0.783 & 0. \\
 0. & 0. & 0.984 \\
\end{array}
\right),\ \ \
U_1^{(D)LR/RL} = 2.087,\ \ \
U_3^{(D)LR/RL} = 1.016.
\end{align}

The QCD-only contribution is
\begin{align}
U^{(q)LL/RR}_{(12)QCD} =
\left(
\begin{array}{cc}
 2.032 & 0 \\
 0 & 0.79 \\
\end{array}
\right),\ \ \
U^{(q)LL/RR}_{(3)QCD} = 1.
\end{align}
\begin{align}
U^{(q)LR/RL}_{(1)QCD} = 2.032,\ \ \
U^{(q)LR/RL}_{(3)QCD} = 1.\ \ \
\end{align}

\subsection{$\Lambda=10\mbox{ TeV}$}

The matrices $U(\mu,\Lambda)$ for $\mu=1\mbox{ GeV}$ and $\Lambda=10\mbox{ TeV}$ are
\begin{align}
U^{(U)LL/RR} = \left(
\begin{array}{ccc}
 2.395 & -0.548 & 0. \\
 -0.009 & 0.748 & 0. \\
 0. & 0. & 1.043 \\
\end{array}
\right),\ \ \
U_1^{(U)LR/RL} = 2.393,\ \ \
U_3^{(U)LR/RL} = 0.958,
\end{align}
\begin{align}
U^{(D)LL/RR} = \left(
\begin{array}{ccc}
 2.369 & 0.273 & 0. \\
 0.004 & 0.75 & 0. \\
 0. & 0. & 0.979 \\
\end{array}
\right),\ \ \
U_1^{(D)LR/RL} = 2.368,\ \ \
U_3^{(D)LR/RL} = 1.022.
\end{align}

The QCD-only contribution is
\begin{align}
U^{(q)LL/RR}_{(12)QCD} =
\left(
\begin{array}{cc}
 2.286 & 0 \\
 0 & 0.759 \\
\end{array}
\right),\ \ \
U^{(q)LL/RR}_{(3)QCD} = 1.
\end{align}
\begin{align}
U^{(q)LR/RL}_{(1)QCD} = 2.286,\ \ \
U^{(q)LR/RL}_{(3)QCD} = 1.\ \ \
\label{eq:ULRRL}   
\end{align}

\section{RGE-evolution Matrices in Non-Chiral Basis}
\label{Sec:App-Non-Chiral}
Here we specify the numerical values of the RGE evolution matrices 
in the P-parity basis. These values are obtained from Eqs.~(\ref{eq:ULLRR})-(\ref{eq:ULRRL})  applying the relations (\ref{eq:op_basis_change-1})-(\ref{eq:op_basis_change-5}).

\subsection{$\Lambda=1\mbox{ TeV}$}
\label{Sec:App-Non-Chiral-1}

The evolution matrices in the $(C^{(q)SS},C^{(q)PP},C^{(q)TT},C^{(q)SP},C^{(q)PS})$ basis, for $u$-type and $d$-type quarks are
\begin{equation}
U^{(U)} = 
\left(
\begin{tikzpicture}[baseline=-.65ex]
\matrix[
  matrix of math nodes,
  column sep=1ex,
] (m)
{
 2.104 & 0. & -0.188 & 0. & 0. \\
 0. & 2.104 & -0.188 & 0. & 0. \\
 -0.006 & -0.006 & 0.781 & 0. & 0. \\
 0. & 0. & 0. & 2.104 & 0. \\
 0. & 0. & 0. & 0. & 2.104 \\
};
\draw[dotted]
  ([xshift=0.5ex]m-1-3.north east) -- ([xshift=3.1ex]m-5-3.south east);
\draw[dotted]
  (m-3-1.south west) -- (m-3-5.south east);
\end{tikzpicture}\mkern-5mu
\right),
\end{equation}
\begin{equation}
U^{(D)} = 
\left(
\begin{tikzpicture}[baseline=-.65ex]
\matrix[
  matrix of math nodes,
  column sep=1ex,
] (m)
{
 2.087 & 0. & 0.094 & 0. & 0. \\
 0. & 2.087 & 0.094 & 0. & 0. \\
 0.003 & 0.003 & 0.783 & 0. & 0. \\
 0. & 0. & 0. & 2.087 & 0. \\
 0. & 0. & 0. & 0. & 2.087 \\
};
\draw[dotted]
  ([xshift=0.5ex]m-1-3.north east) -- ([xshift=2.27ex]m-5-3.south east);
\draw[dotted]
  (m-3-1.south west) -- (m-3-5.south east);
\end{tikzpicture}\mkern-5mu
\right).
\end{equation}
For the isospin singlet and triplet combinations of Wilson Coefficients:
\begin{equation}
C^{(0/3)\Gamma_i\Gamma_j} = C^{(u)\Gamma_i\Gamma_j} \pm C^{(d)\Gamma_i\Gamma_j},
\end{equation}
the $u$- and $d$-type Wilson coeficients mix in each submatrix.
In the
$(C^{(0)SS},C^{(3)SS},C^{(0)PP},C^{(3)PP},C^{(0)TT},C^{(3)TT})$ basis, the evolution matrix reads
\begin{equation}
U^{(0/3)}_{SS,PP,TT} = 
\left(
\begin{tikzpicture}[baseline=-.65ex]
\matrix[
  matrix of math nodes,
  column sep=1ex,
] (m)
{
 2.095 & 0.008 & 0. & 0. & -0.047 & -0.141 \\
 0.008 & 2.095 & 0. & 0. & -0.141 & -0.047 \\
 0. & 0. & 2.095 & 0.008 & -0.047 & -0.141 \\
 0. & 0. & 0.008 & 2.095 & -0.141 & -0.047 \\
 -0.002 & -0.005 & -0.002 & -0.005 & 0.782 & -0.001 \\
 -0.005 & -0.002 & -0.005 & -0.002 & -0.001 & 0.782 \\
};
\end{tikzpicture}\mkern-5mu
\right),
\end{equation}
while for the
$(C^{(0)SP},C^{(3)SP},C^{(0)PS},C^{(3)PS})$ block we have
\begin{equation}
U^{(0/3)}_{SP,PS} = 
\left(
\begin{tikzpicture}[baseline=-.65ex]
\matrix[
  matrix of math nodes,
  column sep=1ex,
] (m)
{
 2.095 & 0.008 & 0. & 0. \\
 0.008 & 2.095 & 0. & 0. \\
 0. & 0. & 2.095 & 0.008 \\
 0. & 0. & 0.008 & 2.095 \\
};
\end{tikzpicture}\mkern-5mu
\right).
\end{equation}
The vector/axial coefficients mix among themselves only, due to the following evolution matrices
\begin{equation}
U^{(U)}_{AA,VV} = U^{(U)}_{VA,AV} = 
\left(
\begin{tikzpicture}[baseline=-.65ex]
\matrix[
  matrix of math nodes,
  column sep=1ex,
] (m)
{
 1. & 0.032 \\
 0.032 & 1. \\
};
\end{tikzpicture}\mkern-5mu
\right),
\end{equation}
\begin{equation}
U^{(D)}_{AA,VV} = U^{(D)}_{VA,AV} = 
\left(
\begin{tikzpicture}[baseline=-.65ex]
\matrix[
  matrix of math nodes,
  column sep=1ex,
] (m)
{
 1. &  -0.016 \\
  -0.016 & 1. \\
};
\end{tikzpicture}\mkern-5mu
\right).
\end{equation}
For the isospin singlet and triplet combinations, in the $(C^{(0)AA/VA},C^{(3)AA/VA},C^{(0)VV/AV},C^{(3)VV/AV})$ basis the evolution matrix is
\begin{equation}
U^{(0/3)}_{AA,VV} = U^{(0/3)}_{VA,AV} = 
\left(
\begin{tikzpicture}[baseline=-.65ex]
\matrix[
  matrix of math nodes,
  column sep=1ex,
] (m)
{
 1. & 0. & 0.008 & 0.024 \\
 0. & 1. & 0.024 & 0.008 \\
 0.008 & 0.024 & 1. & 0. \\
 0.024 & 0.008 & 0. & 1. \\
};
\end{tikzpicture}\mkern-5mu
\right).
\end{equation}

\subsection{$\Lambda=10\mbox{ TeV}$}
\label{Sec:App-Non-Chiral-10}

Here we show the values of the same matrices as in the subsection \ref{Sec:App-Non-Chiral-1}, but for $\Lambda=10$ TeV. 
\begin{equation}
U^{(U)} = 
\left(
\begin{tikzpicture}[baseline=-.65ex]
\matrix[
  matrix of math nodes,
  column sep=1ex,
] (m)
{
 2.394 & 0.001 & -0.274 & 0. & 0. \\
 0.001 & 2.394 & -0.274 & 0. & 0. \\
 -0.009 & -0.009 & 0.748 & 0. & 0. \\
 0. & 0. & 0. & 2.394 & 0.001 \\
 0. & 0. & 0. & 0.001 & 2.394 \\
};
\draw[dotted]
  ([xshift=0.5ex]m-1-3.north east) -- ([xshift=3.2ex]m-5-3.south east);
\draw[dotted]
  (m-3-1.south west) -- (m-3-5.south east);
\end{tikzpicture}\mkern-5mu
\right),
\end{equation}
\begin{equation}
U^{(D)} = 
\left(
\begin{tikzpicture}[baseline=-.65ex]
\matrix[
  matrix of math nodes,
  column sep=1ex,
] (m)
{
 2.369 & 0. & 0.136 & 0. & 0. \\
 0. & 2.369 & 0.136 & 0. & 0. \\
 0.004 & 0.004 & 0.75 & 0. & 0. \\
 0. & 0. & 0. & 2.369 & 0. \\
 0. & 0. & 0. & 0. & 2.369 \\
};
\draw[dotted]
  ([xshift=0.5ex]m-1-3.north east) -- ([xshift=2.3ex]m-5-3.south east);
\draw[dotted]
  (m-3-1.south west) -- (m-3-5.south east);
\end{tikzpicture}\mkern-5mu
\right).
\end{equation}
\begin{equation}
U^{(0/3)}_{SS,PP,TT} = 
\left(
\begin{tikzpicture}[baseline=-.65ex]
\matrix[
  matrix of math nodes,
  column sep=1ex,
] (m)
{
 2.381 & 0.013 & 0. & 0. & -0.069 & -0.205 \\
 0.013 & 2.381 & 0. & 0. & -0.205 & -0.069 \\
 0. & 0. & 2.381 & 0.013 & -0.069 & -0.205 \\
 0. & 0. & 0.013 & 2.381 & -0.205 & -0.069 \\
 -0.002 & -0.007 & -0.002 & -0.007 & 0.749 & -0.001 \\
 -0.007 & -0.002 & -0.007 & -0.002 & -0.001 & 0.749 \\
};
\end{tikzpicture}\mkern-5mu
\right),
\end{equation}
\begin{equation}
U^{(0/3)}_{SP,PS} = 
\left(
\begin{tikzpicture}[baseline=-.65ex]
\matrix[
  matrix of math nodes,
  column sep=1ex,
] (m)
{
 2.381 & 0.013 & 0. & 0. \\
 0.013 & 2.381 & 0. & 0. \\
 0. & 0. & 2.381 & 0.013 \\
 0. & 0. & 0.013 & 2.381 \\
};
\end{tikzpicture}\mkern-5mu
\right).
\end{equation}
\begin{equation}
U^{(U)}_{AA,VV} = U^{(U)}_{VA,AV} = 
\left(
\begin{tikzpicture}[baseline=-.65ex]
\matrix[
  matrix of math nodes,
  column sep=1ex,
] (m)
{
 1.001 & 0.043 \\
 0.043 & 1.001 \\
};
\end{tikzpicture}\mkern-5mu
\right),
\end{equation}
\begin{equation}
U^{(D)}_{AA,VV} = U^{(D)}_{VA,AV} = 
\left(
\begin{tikzpicture}[baseline=-.65ex]
\matrix[
  matrix of math nodes,
  column sep=1ex,
] (m)
{
 1. & -0.021 \\
 -0.021 & 1. \\
};
\end{tikzpicture}\mkern-5mu
\right).
\end{equation}
\begin{equation}
U^{(0/3)}_{AA,VV} = U^{(0/3)}_{VA,AV} = 
\left(
\begin{tikzpicture}[baseline=-.65ex]
\matrix[
  matrix of math nodes,
  column sep=1ex,
] (m)
{
 1. & 0. & 0.011 & 0.032 \\
 0. & 1. & 0.032 & 0.011 \\
 0.011 & 0.032 & 1. & 0. \\
 0.032 & 0.011 & 0. & 1. \\
};
\end{tikzpicture}\mkern-5mu
\right).
\end{equation}

\newpage

\section{Physical Constants}\label{app:constants}

The values of $F_M$ and the dimensionless decay constants ($f_M$) used in the matching conditions (\ref{eq:matchingV})-(\ref{eq:matchingP}) are shown in Table~\ref{tab:decayconstants}.
\begin{table}[tb]
\centering
\begin{tabular}{c|c||c|c}
Decay const. & Value & Decay const. & Value [MeV]\\
\hline\hline
$F_\pi$			& 92.4 MeV	& $F_{K^0}$ & 155.7\\
$F_{\eta_c}$	& 285 MeV   & $F_{D^0}$ & 203.7\\
$f_\rho$		&	0.2		& $F_{B^0}$ & 188\\
$f_\omega$		&	0.059   & $F_{B_s^0}$ & 227.2\\
$f_\phi$		&	0.074   & $F_{K^*}$ & 203.7\\
$f_{J/\psi}$	&	0.134   & $F_{D^*}$ & 223.5 \\
$f_\Upsilon$	&	0.08    & $F_{B^*}$ & 185.9 \\
$f_{f0}$		&	0.28	& $F_{B_s^*}$ & 227.2\\
$f_{a0}$		&	0.19	&\\
\hline\hline
\end{tabular}
\caption{The decays constants used in the matching conditions (\ref{eq:matchingV})-(\ref{eq:matchingP}), taken from Refs.~\cite{Dib:2018rpy} and \cite{Chang:2018aut}.
For the vector meson decay constants, we use experimental values when available, or the values quoted in \cite{Chang:2018aut} from the LQCD model.
}
\label{tab:decayconstants}
\end{table}
For the quark masses, we use the values
\begin{equation}
\tilde m = 7\mbox{ MeV}\,\mbox{\cite{Gasser:1982ap}},\ \ \ 
m_s=25\,\tilde m\,\mbox{\cite{Gasser:1982ap}},\ \ \ 
m_c=1.280\mbox{ GeV}\,\mbox{\cite{Zyla:2020zbs}},\ \ \, m_b=4.18\mbox{ GeV}\, \mbox{\cite{Zyla:2020zbs}}.
\end{equation}
The masses and widths of the mesons considered in this work, taken from Ref.~\cite{Zyla:2020zbs}, are shown in Table~\ref{tab:masses_widths}. When the value quoted in Ref.~\cite{Zyla:2020zbs} corresponds to an interval, we choose the central value.
\begin{table}[tb]
\centering
\begin{tabular}{c|c|c||c|c|c}
Meson & Mass [MeV] & Width & Meson & Mass [MeV] & Width\\
\hline\hline
$\pi^0$	&	134.9768	& 7.7 eV	& $J/\psi$&	3096.9 		& 92.9 keV\\
$\eta$  &	547.862 	& 1.31 keV	& $\Upsilon$&	9460.3 		& 54.02 keV\\
$\eta'$	&	957.78		& 0.188 MeV	& $K^0_L$	& 497.611 & $1.29\times 10^{-8}\mbox{ eV}$ \\
$\eta_c$ &	2983.9		& 32 MeV	& $D^0$ 	& 1864.83 & $1.61\times 10^{-3}\mbox{ eV}$ \\
$f_0(500)$&	475 		& 550 MeV	& $B^0$ 	& 5279.65 & $4.33\times 10^{-4}\mbox{ eV}$ \\
$f_0(980)$& 990 		& 55 MeV	& $B_s^0$	& 5366.88 & $4.31\times 10^{-4}\mbox{ eV}$ \\
$a_0(980)$&	980 		& 75 MeV	& $K^*$ 	& 891.66  & 50.8 MeV \\
$\rho_0$ & 	775.26  	& 149.1 MeV & $D^*$ 	& 2006.85 & - \\
$\omega$ &	782.65  	& 8.49 MeV	& $B^*$ 	& 5324.7 & - \\
$\phi$	&	1019.461 	& 4.249 MeV	& $B_s^*$	& 5415.4 & - \\
\hline\hline
\end{tabular}
\caption{Masses and widths of the mesons considered in this work, taken from Ref.~\cite{Zyla:2020zbs}.}
\label{tab:masses_widths}
\end{table}
For the pseudoscalar mesons $\eta$ and $\eta'$, the singlet-octet mixing angle is $\theta_P=-13.34^{\circ}$~\cite{Ambrosino:2006gk}, with
\begin{eqnarray}
\eta &\to &
-\,\tfrac{1}{\sqrt{2}}\,\sin\delta\, (\bar u u + \bar d d)
-\,\cos\delta\, \bar s s \,,
\nonumber\\
\eta' &\to &
+\,\tfrac{1}{\sqrt{2}}\,\cos\delta\, (\bar u u + \bar d d)
-\,\sin\delta\, \bar s s \,,
\nonumber\\
\delta &=& \theta_P-\theta_I, \qquad \theta_I=\arctan{\tfrac{1}{\sqrt{2}}}\,.
\label{eq:mixing}
\end{eqnarray}

\end{appendix}


\begin{thebibliography}{10}

\bibitem{TheMEG:2016wtm}
A.~M.~Baldini \textit{et al.} [MEG],
Eur. Phys. J. C \textbf{76}, no.8, 434 (2016)
doi:10.1140/epjc/s10052-016-4271-x
[arXiv:1605.05081 [hep-ex]].

\bibitem{Aubert:2009ag}
B.~Aubert \textit{et al.} [BaBar],
Phys. Rev. Lett. \textbf{104}, 021802 (2010)
doi:10.1103/PhysRevLett.104.021802
[arXiv:0908.2381 [hep-ex]].

\bibitem{Hayasaka:2010np}
K.~Hayasaka, K.~Inami, Y.~Miyazaki, K.~Arinstein, V.~Aulchenko, T.~Aushev, A.~M.~Bakich, A.~Bay, K.~Belous and V.~Bhardwaj, \textit{et al.}
Phys. Lett. B \textbf{687}, 139-143 (2010)
doi:10.1016/j.physletb.2010.03.037
[arXiv:1001.3221 [hep-ex]].

\bibitem{Bellgardt:1987du}
U.~Bellgardt \textit{et al.} [SINDRUM],
Nucl. Phys. B \textbf{299}, 1-6 (1988)
doi:10.1016/0550-3213(88)90462-2

\bibitem{Dohmen:1993mp}
C.~Dohmen \textit{et al.} [SINDRUM II],
Phys. Lett. B \textbf{317}, 631-636 (1993)
doi:10.1016/0370-2693(93)91383-X

\bibitem{Aad:2019ugc}
G.~Aad \textit{et al.} [ATLAS],
Phys. Lett. B \textbf{800}, 135069 (2020)
doi:10.1016/j.physletb.2019.135069
[arXiv:1907.06131 [hep-ex]].

\bibitem{Sirunyan:2019shc}
A.~M.~Sirunyan \textit{et al.} [CMS],
JHEP \textbf{03}, 103 (2020)
doi:10.1007/JHEP03(2020)103
[arXiv:1911.10267 [hep-ex]].

\bibitem{CMS:2015hga}
 [CMS],
CMS-PAS-EXO-13-005.

\bibitem{Aad:2020gkd}
G.~Aad \textit{et al.} [ATLAS],
[arXiv:2010.02566 [hep-ex]].

\bibitem{Sirunyan:2018zhy}
A.~M.~Sirunyan \textit{et al.} [CMS],
JHEP \textbf{04}, 073 (2018)
doi:10.1007/JHEP04(2018)073
[arXiv:1802.01122 [hep-ex]].

\bibitem{Aaboud:2018jff}
M.~Aaboud \textit{et al.} [ATLAS],
Phys. Rev. D \textbf{98}, no.9, 092008 (2018)
doi:10.1103/PhysRevD.98.092008
[arXiv:1807.06573 [hep-ex]].

\bibitem{Arndt:2020obb}
K.~Arndt \textit{et al.} [Mu3e],
[arXiv:2009.11690 [physics.ins-det]].

\bibitem{Angelique:2018svf}
J.~C.~Ang\'elique \textit{et al.} [COMET],
[arXiv:1812.07824 [hep-ex]].

\bibitem{Adamov:2018vin}
R.~Abramishvili \textit{et al.} [COMET],
PTEP \textbf{2020}, no.3, 033C01 (2020)
doi:10.1093/ptep/ptz125
[arXiv:1812.09018 [physics.ins-det]].

\bibitem{Miscetti:2020gkk}
S.~Miscetti [Mu2e],
EPJ Web Conf. \textbf{234}, 01010 (2020)
doi:10.1051/epjconf/202023401010

\bibitem{Baldini:2018nnn}
A.~M.~Baldini \textit{et al.} [MEG II],
Eur. Phys. J. C \textbf{78}, no.5, 380 (2018)
doi:10.1140/epjc/s10052-018-5845-6
[arXiv:1801.04688 [physics.ins-det]].

\bibitem{Faessler:2004jt}
A.~Faessler, T.~Gutsche, S.~Kovalenko, V.~E.~Lyubovitskij, I.~Schmidt and F.~Simkovic,
Phys. Lett. B \textbf{590}, 57-62 (2004)
doi:10.1016/j.physletb.2004.03.068
[arXiv:hep-ph/0403033 [hep-ph]].

\bibitem{Faessler:2004ea}
A.~Faessler, T.~Gutsche, S.~Kovalenko, V.~E.~Lyubovitskij, I.~Schmidt and F.~Simkovic,
Phys. Rev. D \textbf{70}, 055008 (2004)
doi:10.1103/PhysRevD.70.055008
[arXiv:hep-ph/0405164 [hep-ph]].

\bibitem{Faessler:2005hx}
A.~Faessler, T.~Gutsche, S.~Kovalenko, V.~E.~Lyubovitskij and I.~Schmidt,
Phys. Rev. D \textbf{72}, 075006 (2005)
doi:10.1103/PhysRevD.72.075006
[arXiv:hep-ph/0507033 [hep-ph]].

\bibitem{Dib:2018rpy}
C.~O.~Dib, T.~Gutsche, S.~G.~Kovalenko, V.~E.~Lyubovitskij and I.~Schmidt,
Phys. Rev. D \textbf{99}, no.3, 035020 (2019)
doi:10.1103/PhysRevD.99.035020
[arXiv:1812.02638 [hep-ph]].

\bibitem{Gninenko:2018num}
S.~Gninenko, S.~Kovalenko, S.~Kuleshov, V.~E.~Lyubovitskij and A.~S.~Zhevlakov,
Phys. Rev. D \textbf{98}, no.1, 015007 (2018)
doi:10.1103/PhysRevD.98.015007
[arXiv:1804.05550 [hep-ph]].

\bibitem{Nussinov:2000nm}
S.~Nussinov, R.~D.~Peccei and X.~M.~Zhang,
Phys. Rev. D \textbf{63}, 016003 (2001)
doi:10.1103/PhysRevD.63.016003
[arXiv:hep-ph/0004153 [hep-ph]].

\bibitem{Gutsche:2011bi}
T.~Gutsche, J.~C.~Helo, S.~Kovalenko and V.~E.~Lyubovitskij,
Phys. Rev. D \textbf{83}, 115015 (2011)
doi:10.1103/PhysRevD.83.115015
[arXiv:1103.1317 [hep-ph]].

\bibitem{Ambrose:1998us}
D.~Ambrose \textit{et al.} [BNL],
Phys. Rev. Lett. \textbf{81}, 5734-5737 (1998)
doi:10.1103/PhysRevLett.81.5734
[arXiv:hep-ex/9811038 [hep-ex]].

\bibitem{Aaij:2015qmj}
R.~Aaij \textit{et al.} [LHCb],
Phys. Lett. B \textbf{754}, 167-175 (2016)
doi:10.1016/j.physletb.2016.01.029
[arXiv:1512.00322 [hep-ex]].

\bibitem{Aaij:2017cza}
R.~Aaij \textit{et al.} [LHCb],
JHEP \textbf{03}, 078 (2018)
doi:10.1007/JHEP03(2018)078
[arXiv:1710.04111 [hep-ex]].

\bibitem{Aubert:2008cu}
B.~Aubert \textit{et al.} [BaBar],
Phys. Rev. D \textbf{77}, 091104 (2008)
doi:10.1103/PhysRevD.77.091104
[arXiv:0801.0697 [hep-ex]].

\bibitem{Aaij:2019okb}
R.~Aaij \textit{et al.} [LHCb],
Phys. Rev. Lett. \textbf{123}, no.21, 211801 (2019)
doi:10.1103/PhysRevLett.123.211801
[arXiv:1905.06614 [hep-ex]].

\bibitem{Gonzalez:2015ady}
M.~Gonz\'alez, M.~Hirsch and S.~G.~Kovalenko,
Phys. Rev. D \textbf{93}, no.1, 013017 (2016)
[erratum: Phys. Rev. D \textbf{97}, no.9, 099907 (2018)]
doi:10.1103/PhysRevD.93.013017
[arXiv:1511.03945 [hep-ph]].

\bibitem{Gonzalez:2017mcg}
M.~Gonz\'alez, M.~Hirsch and S.~Kovalenko,
Phys. Rev. D \textbf{97}, no.11, 115005 (2018)
doi:10.1103/PhysRevD.97.115005
[arXiv:1711.08311 [hep-ph]].

\bibitem{Arbelaez:2016uto}
C.~Arbel\'aez, M.~Gonz\'alez, S.~Kovalenko and M.~Hirsch,
Phys. Rev. D \textbf{96}, no.1, 015010 (2017)
doi:10.1103/PhysRevD.96.015010
[arXiv:1611.06095 [hep-ph]].

\bibitem{Arbelaez:2016zlt}
C.~Arbel\'aez, M.~Gonz\'alez, M.~Hirsch and S.~Kovalenko,
Phys. Rev. D \textbf{94}, no.9, 096014 (2016)
[erratum: Phys. Rev. D \textbf{97}, no.9, 099904 (2018)]
doi:10.1103/PhysRevD.94.096014
[arXiv:1610.04096 [hep-ph]].

\bibitem{Buras:1998raa}
A.~J.~Buras,
[arXiv:hep-ph/9806471 [hep-ph]].

\bibitem{Davidson:2016edt}
S.~Davidson,
Eur. Phys. J. C \textbf{76}, no.7, 370 (2016)
doi:10.1140/epjc/s10052-016-4207-5
[arXiv:1601.07166 [hep-ph]].

\bibitem{Jenkins:2017dyc}
E.~E.~Jenkins, A.~V.~Manohar and P.~Stoffer,
JHEP \textbf{01}, 084 (2018)
doi:10.1007/JHEP01(2018)084
[arXiv:1711.05270 [hep-ph]].

\bibitem{Buchalla:1989we}
G.~Buchalla, A.~J.~Buras and M.~K.~Harlander,
Nucl. Phys. B \textbf{337}, 313-362 (1990)
doi:10.1016/0550-3213(90)90275-I

\bibitem{Buras:1993dy}
A.~J.~Buras, M.~Jamin and M.~E.~Lautenbacher,
Nucl. Phys. B \textbf{408}, 209-285 (1993)
doi:10.1016/0550-3213(93)90535-W
[arXiv:hep-ph/9303284 [hep-ph]].

\bibitem{Aebischer:2017gaw}
J.~Aebischer, M.~Fael, C.~Greub and J.~Virto,
JHEP \textbf{09}, 158 (2017)
doi:10.1007/JHEP09(2017)158
[arXiv:1704.06639 [hep-ph]].

\bibitem{Chang:2018aut}
Q.~Chang, X.~N.~Li, X.~Q.~Li and F.~Su,
Chin. Phys. C \textbf{42}, no.7, 073102 (2018)
doi:10.1088/1674-1137/42/7/073102
[arXiv:1805.00718 [hep-ph]].

\bibitem{Gasser:1982ap}
J.~Gasser and H.~Leutwyler,
Phys. Rept. \textbf{87}, 77-169 (1982)
doi:10.1016/0370-1573(82)90035-7

\bibitem{Zyla:2020zbs}
P.~A.~Zyla \textit{et al.} [Particle Data Group],
PTEP \textbf{2020}, no.8, 083C01 (2020)
doi:10.1093/ptep/ptaa104

\bibitem{Ambrosino:2006gk}
F.~Ambrosino \textit{et al.} [KLOE],
Phys. Lett. B \textbf{648}, 267-273 (2007)
doi:10.1016/j.physletb.2007.03.032
[arXiv:hep-ex/0612029 [hep-ex]].

\end{thebibliography}
\end{document}